\documentclass[iop]{emulateapj}  

\usepackage{color}
\usepackage{hyperref}  
\usepackage{breakurl}  

\newcommand\kms{\ifmmode{\rm km\thinspace s^{-1}}\else km\thinspace s$^{-1}$\fi}

\shortauthors{Torres}
\shorttitle{Rapidly rotating stars in the Pleiades}

\begin{document} 
\submitted{Accepted for publication in The Astrophysical Journal}

\title{Spectroscopic monitoring of rapidly-rotating early-type stars in the
  Pleiades cluster}

\author{Guillermo Torres}

\affiliation{Center for Astrophysics $\vert$ Harvard \& Smithsonian,
  60 Garden St., Cambridge, MA 02138, USA; gtorres@cfa.harvard.edu}

\begin{abstract} 

Radial-velocities for the early-type stars in the Pleiades cluster have always been challenging to measure because of the significant rotational broadening of the spectral lines. The large scatter in published velocities has led to claims that many are spectroscopic binaries, and in several cases preliminary orbital solutions have been proposed.  To investigate these claims, we obtained and report here velocity measurements for 33 rapidly-rotating B, A, and early F stars in the Pleiades region, improving significantly on the precision of the historical velocities for most objects.  With one or two exceptions, we do not confirm any of the previous claims of variability, and we also rule out all four of the previously published orbital solutions, for HD~22637, HD~23302, HD~23338, and HD~23410.  We do find HD~22637 to be a binary, but with a different period (71.8~days). HD~23338 is likely a binary as well, with a preliminary 8.7~yr period also different from the one published.  Additionally, we report a 3635-day orbit for HD~24899, another new spectroscopic binary in the cluster. From the 32 bona fide members in our sample we determine a mean radial velocity for the Pleiades of $5.79\pm0.24\,\kms$, or $5.52\pm0.31\,\kms$ when objects with known visual companions are excluded.  Adding these astrometric binaries to the new spectroscopic ones, we find a lower limit to the binary fraction among the B and A stars of 37\%. In addition to the velocities, we measure $v\sin i$ for all stars, ranging between 69 and 317~\kms.
\end{abstract}

\section{Introduction}
\label{sec:introduction}

The Pleiades cluster has been the subject of numerous spectroscopic
surveys of its brighter members going back more than a century
\citep[e.g.,][]{Adams:1904, Frost:1926, Smith:1944, Abt:1965,
  Pearce:1975, Liu:1991, Mermilliod:2009}.  One of the most extensive
programs, summarized in the last of these references, focused on
slowly rotating FGK stars and ran for nearly 20 years, leading to the
discovery and orbit characterization of many new binary systems in the
cluster \citep{Mermilliod:1992, Mermilliod:1997, Rosvick:1992,
  Raboud:1998}.  While these later-type stars typically offer no
difficulty for the determination of their radial velocities (RVs),
earlier stars of spectral type B and A are much more challenging. Not
only do they display fewer lines in their spectra, but the features
are typically very broad because these stars tend to be rotating very
rapidly, some with projected rotational velocities $v \sin i$ in
excess of 200~\kms. As a result, available RVs for B and A stars in
the Pleiades, and for a few F stars, are of relatively poor quality
and often show significant scatter.  Over the years more than a dozen
of these objects have been claimed to be possible spectroscopic
binaries, and in some cases tentative orbits have been published, but
few have ever been confirmed.

A main motivation for this paper is therefore to investigate some of
these claims, based on new observations gathered in the course of a
long-term spectroscopic survey of the Pleiades ongoing at the author's
institution. The ultimate goal of that survey is to achieve a more
complete census of the spectroscopic binaries in the cluster,
extending it to longer-period systems in the regime where
spectroscopic and astrometric techniques overlap, thereby enabling the
determination of absolute masses.  In addition to the many later-type
stars whose velocities can be derived using standard cross-correlation
techniques, the target list in the Pleiades includes nearly three
dozen rapidly rotating B, A, and early F stars, which experience has
shown require a different approach in order to derive accurate RVs.
These objects therefore separate themselves naturally from the rest of
the survey, and are the subject of this paper. A further motivation
for this work is to derive accurate rotational velocities for all the
fast rotators in a uniform way.

We begin by defining the sample in Section~\ref{sec:sample}, which we
follow with a description in Section~\ref{sec:observations} of our
observations and the methodology for the determination of radial
velocities.  The same section explains how we infer the rotational
broadening for each star.  Next we present the results of our RV
determinations (Section~\ref{sec:rvresults}), including new
spectroscopic orbits for three of the objects, and detailed notes for
many of the others. This section also includes a discussion of the
mean radial velocity of the cluster. Our new rotational velocity
measurements are then reported in Section~\ref{sec:rotresults}. Final
thoughts are presented in Section~\ref{sec:conclusions}.

\section{Sample}
\label{sec:sample}

The B, A, and early F stars discussed here are part of a larger survey
based initially on an unpublished list of stars in the Pleiades area,
that was maintained at the author's institution by John Stauffer and
Charles Prosser, and was originally drawn from classical sources of
common proper motion members \citep{Hertzsprung:1947, Artyukhina:1970,
  Jones:1981, Haro:1982, vanLeeuwen:1986}.  This compilation was later
augmented with probable members from other published lists, mostly
fainter than the original set. For this work we focus on the 33
bright, early type stars that are rotating rapidly, and for which
standard cross-correlation techniques for RV determination produce
poor results. We define poor results here as those with average formal
uncertainties larger than about 3~\kms\ for a given star (see below),
which can make it difficult to detect variability in long-period
orbits.  This will typically be the case for stars rotating more
rapidly than $v \sin i \approx 100~\kms$, or for stars rotating
slightly less rapidly but that are earlier than about spectral type
A0, and therefore have fewer lines.

\begin{figure}
\epsscale{1.15}
\plotone{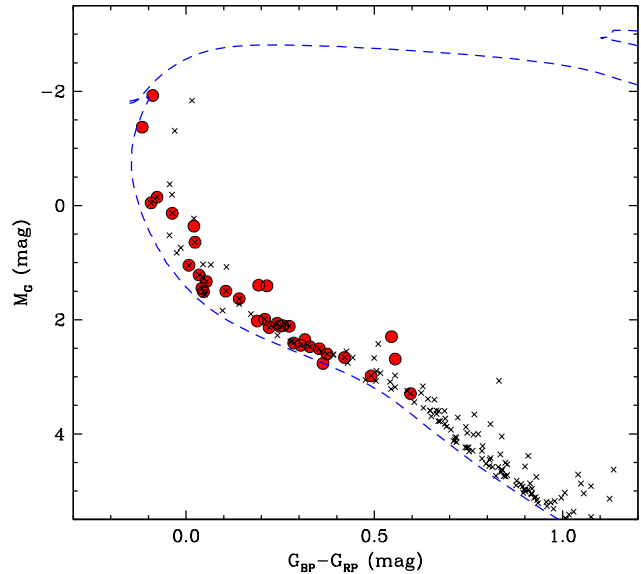}

\figcaption{Location of the rapidly rotating B, A, and F stars
  observed for this project in a diagram of absolute $G$ magnitude
  versus $G_{\rm BP}-G_{\rm RP}$ color for the Pleiades (red circles).
  Probable kinematic members from the list of \cite{Gao:2019} are
  represented with crosses. Ones in the upper main-sequence that are
  not also marked with a red circle are either stars that are not
  rapid rotators, or stars we did not observe.  Conversely, red
  circles without crosses are either stars with poor {\it Gaia\/}
  astrometry (e.g., too bright) that are therefore not included in the
  list of \cite{Gao:2019}, or else non-members.  Absolute magnitudes
  were computed using the individual parallax of each star. In a few
  cases in which the parallax is uncertain, we have adopted a value
  derived from the proper motion, following equations 7 and 8 of
  \cite{Gao:2019}.  Also shown for reference is a model isochrone from
  the PARSEC series by \cite{Chen:2014} for solar metallicity and an
  age of 125~Myr, with extinction and reddening applied based on an
  adopted average value of $E(B-V) = 0.04$ for the cluster. The
  isochrone is seen not to provide a good match to the
  observations.\label{fig:cmd}}

\end{figure}

The 33 rapidly rotating B, A, and F stars in our sample are listed in
Table~\ref{tab:sample}, which includes other common names used among
Pleiades members, and spectral types as listed on SIMBAD.  The
location of the targets in the color-magnitude diagram of the cluster
is indicated in Figure~\ref{fig:cmd}, showing other objects from the
membership list of \cite{Gao:2019}.

\setlength{\tabcolsep}{3pt}  
\begin{deluxetable}{llccl}
\tablecaption{Stars in Our Sample \label{tab:sample}}
\tablehead{
\colhead{Name} &
\colhead{Other Name} &
\colhead{R.A.} &
\colhead{Dec.} &
\colhead{SpT}
}
\startdata
 HD 21744  &   AK III-153  &  03 31 15.95  &  +25 15 19.7  &  A3      \\ 
 HD 22578  &   TRU S23     &  03 38 40.72  &  +22 39 34.6  &  A0      \\ 
 HD 22614  &   TRU S25     &  03 39 06.72  &  +24 42 10.3  &  A0      \\ 
 HD 22637  &   TRU S26     &  03 39 13.20  &  +21 50 35.7  &  A0V     \\ 
 HD 22702  &   TRU S29     &  03 39 51.16  &  +25 11 41.5  &  F1IV    \\ 
 HD 23155  &   HII 153     &  03 43 43.20  &  +25 04 50.6  &  A2      \\ 
 HD 23302  &   HII 468     &  03 44 52.53  &  +24 06 48.1  &  B6IIIe  \\ 
 HD 23324  &   HII 541     &  03 45 09.74  &  +24 50 21.3  &  B8V     \\ 
 HD 23338  &   HII 563     &  03 45 12.50  &  +24 28 02.2  &  B6IV    \\ 
 HD 23323  &   AK III-909  &  03 45 15.01  &  +26 53 29.1  &  A5      \\ 
 HD 23361  &   HII 652     &  03 45 26.14  &  +24 02 06.6  &  A3V     \\ 
 HD 23388  &   TRU S76     &  03 45 31.99  &  +21 14 48.1  &  A3      \\ 
 HD 23402  &   TRU S78     &  03 45 39.90  &  +22 41 40.1  &  A0      \\ 
 HD 23410  &   HII 801     &  03 45 48.81  &  +23 08 49.7  &  A0Va    \\ 
 HD 23409  &   HII 804     &  03 45 51.64  &  +24 02 20.0  &  A2V     \\ 
 HD 23432  &   HII 817     &  03 45 54.48  &  +24 33 16.2  &  B8V     \\ 
 HD 23430  &   TRU S84     &  03 45 59.14  &  +25 23 54.9  &  A0      \\ 
 HD 23489  &   HII 1028    &  03 46 27.28  &  +24 15 18.0  &  A2V     \\ 
 HD 23512  &   HII 1084    &  03 46 34.19  &  +23 37 26.5  &  A2V     \\ 
 HD 23585  &   HII 1284    &  03 47 04.21  &  +23 59 42.8  &  F0V     \\ 
 HD 23629  &   HII 1375    &  03 47 21.04  &  +24 06 58.6  &  A0V     \\ 
 HD 23643  &   HII 1425    &  03 47 26.83  &  +23 40 42.0  &  A3Van   \\ 
 HD 23753  &   HII 1823    &  03 48 20.82  &  +23 25 16.5  &  B8V     \\ 
 HD 23763  &   HII 1876    &  03 48 30.10  &  +24 20 43.8  &  A2V     \\ 
 HD 23852  &   TRU S137    &  03 49 11.26  &  +22 36 34.1  &  A0      \\ 
 HD 23863  &   HII 2195    &  03 49 12.19  &  +23 53 12.4  &  A7Vn    \\ 
 HD 23912  &   HII 2345    &  03 49 32.72  &  +23 22 49.4  &  F4Vn    \\ 
 HD 23913  &   TRU S142    &  03 49 38.18  &  +22 32 00.5  &  B9      \\ 
 HD 23950  &   TRU S149    &  03 49 55.07  &  +22 14 39.0  &  B8III   \\ 
 HD 24013  &   HII 2690    &  03 50 28.05  &  +24 29 43.7  &  A2      \\ 
 HD 24178  &   TRU S165    &  03 51 57.42  &  +25 59 55.9  &  A0      \\ 
 HD 24711  &   TRU S185    &  03 56 28.12  &  +23 09 01.0  &  A0      \\ 
 HD 24899  &   TRU S194    &  03 58 20.90  &  +24 04 52.0  &  B9V        
\enddata

\tablecomments{The ``AK'', ``HII'', and ``TRU'' nomenclature
  originate from \cite{Artyukhina:1970}, \cite{Hertzsprung:1947}, and
  \cite{Trumpler:1921}, respectively. ICRS coordinates are taken from
  the {\it Gaia}/DR2 catalog \citep{Gaia:2018a}, and spectral types are
  from SIMBAD.}

\end{deluxetable}
\setlength{\tabcolsep}{6pt}  

\section{Observations and methods}
\label{sec:observations}

Spectroscopic monitoring of the B, A, and F stars in our sample was
carried out between 2009 October and 2020 January with the Tillinghast
Reflector Echelle Spectrograph \citep[TRES;][]{Szentgyorgyi:2007,
  Furesz:2008}. This is a bench-mounted, fiber-fed instrument attached
to the 1.5\,m Tillinghast Reflector at the Fred L.\ Whipple
Observatory on Mount Hopkins (AZ). It delivers a resolving power of $R
\approx 44,\!000$, and covers the wavelength region
3800--9100\,\AA\ in 51 orders. We collected a total of 388 spectra for
our list of 33 stars. The signal-to-noise ratios at a mean wavelength
of 5187\,\AA\ vary considerably, both for a given star and also
between stars, and range from 30 to about 600 per resolution element
of 6.8~\kms. A handful of spectra had exposures that were too weak to
be useful, and were excluded.  Reductions were performed with a
dedicated pipeline that applies standard dark, bias, and flatfield
corrections, and uses the three back-to-back exposures typically
obtained for each object to aid in removing cosmic rays before
combining them. The wavelength calibration was based on exposures of a
thorium-argon lamp taken before and after each science exposure.

Careful examination of the line profiles showed no convincing evidence
of asymmetries attributable to double lines, which, as indicated
later, have been reported previously for a few of these stars. We
therefore proceeded on the assumption that they are all effectively
single-lined.  We note also that none of the stars display Balmer line
emission.

Radial-velocity determinations were first attempted using the same
standard cross-correlation techniques that are very effective for
solar-type stars with relatively sharp lines, such as those making up
the bulk of the objects in the larger Pleiades survey. Our approach
for those stars uses synthetic templates from a large pre-computed
library based on model atmospheres by R.\ L.\ Kurucz, and a line list
tuned to better match the spectra of real stars
\citep[see][]{Nordstrom:1994, Latham:2002}. The templates cover a
wavelength window of $\sim$300\,\AA\ centered around 5187\,\AA, which
contains the \ion{Mg}{1}\,b triplet as well as numerous iron lines.
This region (particularly the $\sim$100\,\AA\ order centered on the Mg
triplet) has been found from experience to provide the best
information for the measurement of RVs, usually yielding results with
internal errors of 0.5\,\kms\ or less for stars with modest rotation.
However, for the early-type stars that constitute about 10\% of the
survey, the iron lines are no longer visible and the \ion{Mg}{1}
triplet is much weaker, particularly if the stars are rotating
rapidly. The velocity measurements then become very difficult: they
have individual formal uncertainties as large as 5--10\,\kms, or
sometimes more, and result in a large scatter for any given object.

Methodologies involving cross-correlation with CCD detectors for
rapidly-rotating early-type stars have been investigated, e.g., by
\cite{Morse:1991}. Those authors identified two regions about
140\,\AA\ wide that seemed most promising at their spectral resolution
of 44\,\kms. One is centered around 3787\,\AA, but is outside the
range covered by our spectra. The other is centered around 4073\,\AA,
and features H$\delta$ as well as several \ion{He}{1} lines. This
happens to be near the middle of one of our spectral orders. We
explored the use of this region by creating synthetic spectra with the
{\tt SPECTRUM\/} code of \cite{Gray:1994}, using Kurucz model
atmospheres for a range of temperatures appropriate for B, A, and
early F stars, along with the line list provided with the program.
However, we still found the velocities to be poor, perhaps because the
order centered at 4073\,\AA\ in our spectra is only 70\,\AA\ wide,
half of what \cite{Morse:1991} used. We also investigated the use of
an order containing the \ion{He}{1} $\lambda$4471 and \ion{Mg}{2}
$\lambda$4481 lines (the latter being a blend of two Mg lines), which
are strong in hot stars, but we obtained similarly disappointing
results.

Least-squares deconvolution (LSD) is another technique with somewhat
similar benefits as cross-correlation, in the sense that it can use
all of the information available over a wide wavelength range, and
delivers a mean line profile for the star with a higher
signal-to-noise ratio than the individual spectra \citep[see,
  e.g.,][]{Kochukhov:2010}. However, our attempts with LSD did not
result in much improvement compared to cross-correlation, giving only
marginally smaller uncertainties for the radial velocities.

In the end we opted for a more classical approach, and measured the
centroids of five strong Balmer lines manually, by fitting a Gaussian
curve to the core of each line profile using the {\tt splot} task
within IRAF.\footnote{IRAF is distributed by the National Optical
  Astronomy Observatories, which is operated by the Association of
  Universities for Research in Astronomy, Inc., under contract with
  the National Science Foundation.} We consider this procedure to be
adequate for these stars given that all of the spectra appear
effectively single-lined, as mentioned earlier.  The five lines
considered are H$\alpha$, H$\beta$, H$\gamma$, H$\delta$, and
H$\zeta$. The H$\beta$ and H$\delta$ lines are each present in two
adjacent orders, and we measured both independently.  The wavelengths
adopted for these lines were taken from the NIST Atomic Spectra
Database.\footnote{\url{https://physics.nist.gov/PhysRefData/ASD/lines_form.html}}
Table~\ref{tab:rvs} collects the individual velocities derived for
each star, calculated from the straight average of the 7 line
measurements from each spectrum, with internal uncertainties given by
the error of the mean. The average internal uncertainty for the 388
velocities is 1.7\,\kms, and most errors are under 3\,\kms.

\begin{deluxetable}{lccc}
\tablecaption{Heliocentric Radial Velocity Measurements \label{tab:rvs}}
\tablehead{
\colhead{} &
\colhead{HJD} &
\colhead{RV} &
\colhead{$\sigma_{\rm RV}$}
\\
\colhead{Name} &
\colhead{(2,400,000+)} &
\colhead{(\kms)} &
\colhead{(\kms)}
}
\startdata
HD 21744  &  58175.6005  &  4.66  &  2.04 \\
HD 21744  &  58187.6059  &  5.69  &  2.41 \\
HD 21744  &  58205.6077  &  8.28  &  2.89 \\
HD 21744  &  58409.8519  &  2.75  &  1.83 \\
HD 21744  &  58442.6960  &  5.39  &  1.75 
\enddata
\tablecomments{(This table is available in its entirety in machine-readable form.)}
\end{deluxetable}

\begin{figure}
\epsscale{1.15}
\begin{tabular}{cc}
\includegraphics[width=4.0cm]{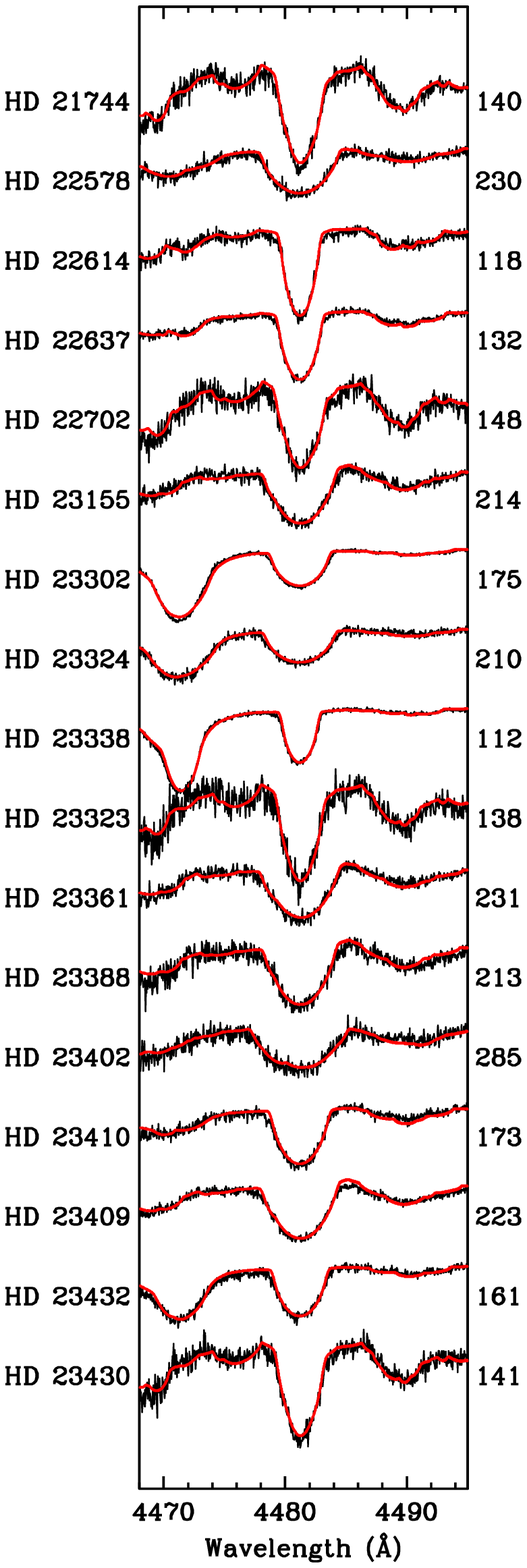} &
\includegraphics[width=4.0cm]{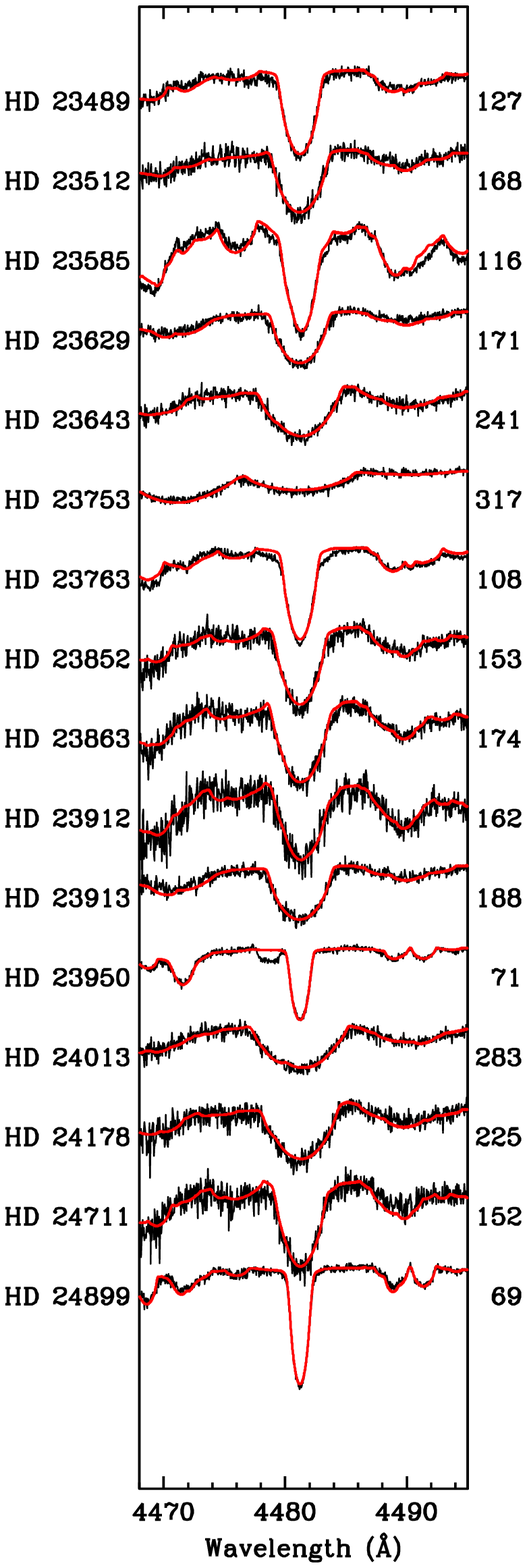}
\end{tabular}

\figcaption{Coadded spectrum of each star centered on the \ion{Mg}{2}
  $\lambda$4481 line, compared against the corresponding synthetic
  spectrum (smooth red line) for the measured $v \sin i$, labeled on
  the right side of each panel.\label{fig:coadd}}

\end{figure}

In addition to measuring the velocities, we made a determination of
the rotational broadening for the stars in our sample. One of them
(HD~23323) appears to have no previous determination of $v \sin i$ in
the literature, while others have sometimes wildly discrepant values
from different sources. We cross-correlated each spectrum against
synthetic templates calculated with the {\tt SPECTRUM\/} code for a
range of rotational broadenings up to 400\,\kms. We adopted a fixed
temperature for each star appropriate for its spectral type, as we
found that the $v \sin i$ measurements are insensitive to that
value. We explored several regions of the spectrum including the one
featuring the \ion{Mg}{1}\,b triplet, and an order containing the
\ion{He}{1} $\lambda$4471 and \ion{Mg}{2} $\lambda$4481 lines, which
are commonly used for this purpose. Even though the order with the
latter lines did not yield useful velocities, as we indicated earlier,
we found that it performed best for determining $v \sin i$, giving
fairly consistent answers from repeat observations of the same star.
To be conservative, we adopted the standard deviation of the
rotational broadenings as a measure of their uncertainty, rather than
the error of the mean. To illustrate these results,
Figure~\ref{fig:coadd} shows the coadded spectra for each star in the
region of the \ion{Mg}{2} $\lambda$4481 line, compared against
synthetic spectra with the measured $v \sin i$ values, as labeled.

\setlength{\tabcolsep}{4pt}  
\setlength{\tabcolsep}{6pt}  

\section{Radial velocity results}
\label{sec:rvresults}

\setlength{\tabcolsep}{4pt}  
\begin{deluxetable*}{lccccccccl}
\tablecaption{Summary of Radial Velocity Results \label{tab:rvsummary}}
\tablehead{
\colhead{} &
\colhead{} &
\colhead{} &
\colhead{HJD} &
\colhead{Span} &
\colhead{$\langle RV\rangle$} &
\colhead{$\sigma_{\rm RV}$} &
\colhead{$RV_{\rm pred}$} &
\colhead{} &
\colhead{}
\\
\colhead{Name} &
\colhead{$N_{\rm obs}$} &
\colhead{$\langle$S/N$\rangle$} &
\colhead{(2,400,000+)} &
\colhead{(days)} &
\colhead{(\kms)} &
\colhead{(\kms)} &
\colhead{(\kms)} &
\colhead{Binarity} &
\colhead{RV Variability}
}
\startdata
 HD 21744   &  11  &  78  &  58176--58887  &   711  &   4.59  &   1.75  &   4.23   & 2\farcs2, 6.1 (1)      & \\
 HD 22578   &   8  & 162  &  58176--58895  &   719  &   7.91  &   1.00  &   6.07   &                        & \\
 HD 22614 * &   9  & 127  &  58176--58766  &   590  &   5.35  &   0.91  &   5.02   &                        & var(S) \\
 HD 22637 * &  32  & 125  &  58136--58879  &   743  &   3.94  &   0.78  &   6.53   &                        & orb(P1); New orbit \\
 HD 22702 * &  10  &  61  &  58176--58895  &   719  &   4.58  &   1.55  &   4.80   &                        & \\
 HD 23155 * &  10  & 110  &  58177--58766  &   589  &   5.07  &   1.47  &   5.07   &                        & var(S) \\
 HD 23302 * &  31  & 352  &  58081--58880  &   799  &   6.98  &   1.00  &   5.64   & 0\farcs2*, 3.6 (2)     & orb(A,J), var(H,P,P1,L) \\
 HD 23324 * &   9  & 227  &  58175--58769  &   594  &   7.39  &   1.72  &   5.27   & 0\farcs05*, 0.0 (2)    & var(A,H,P1,L) \\
 HD 23338 * &  14  & 308  &  58174--58854  &   680  &   5.66  &   0.34  &   5.47   & 0\farcs19, 2.3 (3)     & orb(A), var(P1,L); New orbit \\
 HD 23323 * &   7  &  54  &  55166--56937  &  1771  &   3.78  &   1.49  &   4.20   &                        & \\
 HD 23361 * &  10  & 132  &  56694--58769  &  2075  &   4.65  &   1.89  &   5.71   &                        & var(P1,L) \\
 HD 23388   &   9  &  89  &  58177--58749  &   572  &   4.81  &   1.63  &   7.20   &                        & \\
 HD 23402   &   8  &  98  &  58177--58505  &   328  &   4.57  &   2.80  &   6.43   &                        & \\
 HD 23410 * &  10  & 144  &  58177--58744  &   567  &   6.37  &   1.69  &   6.20   & 3\farcs5, 3.2 (1)      & orb(A), var(P1) \\
 HD 23409 * &  19  & 135  &  56294--58478  &  2184  &   4.05  &   1.81  &   5.73   &                        & var(L) \\
 HD 23432   &  11  & 188  &  58175--58875  &   700  &   6.59  &   1.06  &   5.46   &                        & \\
 HD 23430   &   8  &  72  &  58177--58771  &   594  &   4.73  &   0.84  &   5.02   &                        & \\  
 HD 23489 * &  12  & 116  &  55172--57670  &  2498  &   6.20  &   0.84  &   5.65   & 0\farcs33 (4)          & var(L) \\
 HD 23512 * &   9  &  84  &  58177--58879  &   702  &   7.69  &   1.37  &   5.99   & 0\farcs1*, 2.0 (1)     & var(S,P1,L) \\
 HD 23585 * &  24  &  96  &  55172--58505  &  3333  &   5.70  &   1.07  &   5.82   &                        & var(L) \\
 HD 23629 * &   9  & 155  &  58175--58771  &   596  &   5.75  &   1.91  &   5.77   & 0\farcs0019*, 1.1 (5)  & var(S,P1,L) \\
 HD 23643 * &   8  & 111  &  58177--58771  &   594  &   5.96  &   1.73  &   6.00   &                        & var(L) \\
 HD 23753 * &   9  & 247  &  58175--58552  &   377  &   7.07  &   2.32  &   6.19   & 0\farcs10*, 2.0 (1)    & var(F,P1) \\
 HD 23763 * &  23  & 136  &  55172--58452  &  3280  &   7.32  &   1.83  &   5.70   & 0\farcs2*, 1.3 (1)     & var(A,P1,L) \\
 HD 23852   &   8  &  91  &  58177--58887  &   710  &   6.31  &   1.20  &   6.67   &                        & \\
 HD 23863 * &   6  &  95  &  55589--58419  &  2830  &   4.58  &   1.02  &   5.98   & 0\farcs02*, 4.1 (6)    & var(A,P1,L) \\
 HD 23912   &   5  &  64  &  58412--58771  &   359  &   5.40  &   0.67  &   6.27   &                        & \\
 HD 23913   &   7  & 137  &  58177--58771  &   594  &   6.76  &   1.37  &   6.73   &                        & \\
 HD 23950 * &  13  & 238  &  55589--58466  &  2877  &   9.41  &   0.48  &   6.90   &                        & var(H) \\
 HD 24013 * &  11  & 111  &  58177--58837  &   660  &   9.41  &   2.48  & \nodata  & 0\farcs0155*, 1.1 (7)  & var(L); Non-member \\
 HD 24178   &   7  &  92  &  58419--58763  &   344  &   3.87  &   1.29  &   5.00   &                        & \\
 HD 24711 * &   5  &  76  &  58177--58501  &   324  &   5.20  &   1.38  &   6.75   &                        & var(S) \\
 HD 24899 * &  16  & 135  &  55107--58501  &  3394  &   7.18  &   0.07  &   6.34   & (8)                    & New orbit
\enddata

\tablecomments{Asterisks after the names call attention to notes in
  the text. Subsequent columns give the number of observations, the
  average signal-to-noise ratio per resolution element, the Julian day
  interval, and the time span. Next we list the average radial
  velocity, $\langle RV\rangle$, and the standard deviation of the
  velocities for each star, $\sigma_{\rm RV}$, which we adopt here as
  the error of the mean velocity, to be conservative. For HD~22637,
  HD~23338, and HD~24899, the velocities and uncertainties listed
  correspond to the center of mass velocity of the binary, based on
  the new orbital solutions in this paper (Table~\ref{tab:orbits} and
  Table~\ref{tab:elem_HD23338}). $RV_{\rm pred}$ is the predicted
  radial velocity within the cluster, based on the position of the
  convergent point and the mean distance and proper motion of the
  Pleiades from the {\it Gaia\/} mission \citep{Gaia:2018b}. It is not
  given for HD~24013, which is a background star and not a cluster
  member. Uncertainties for the predicted velocities are very small
  ($< 0.1$~\kms). The Binarity column identifies stars reported to be
  close astrometric binaries, and gives the angular separation and
  magnitude difference of the pair (when available). An asterisk
  following the separation indicates the measurement is from a lunar
  occultation event, in which case it corresponds strictly to a
  separation projected in the direction of the lunar motion. Codes in
  parentheses refer to the following sources: (1) Washington Double
  Star Catalog \citep{Mason:2001}; (2) \cite{Richichi:2002}; (3)
  \cite{Richichi:1994}; (4) \cite{Guerrero:2020}; (5)
  \cite{McGraw:1974}; (6) \cite{Richichi:2012}; (7) \cite{Qian:1991};
  (8) \cite{ESA:1997}.  The last column indicates previous claims of
  velocity variability from the literature (``var''), or published
  orbital solutions (``orb''). Codes in parentheses correspond to the
  following sources: (F) \cite{Frost:1926}; (S) \cite{Smith:1944}; (A)
  \cite{Abt:1965}; (H) \cite{Hube:1970}; (P) \cite{Pearce:1971}; (P1)
  \cite{Pearce:1975}; (J) \cite{Jarad:1989}; and (L) \cite{Liu:1991}.
  The three objects with new orbits from this work are also noted.}

\end{deluxetable*}
\setlength{\tabcolsep}{6pt}  

Table~\ref{tab:rvsummary} summarizes the results of our RV
determinations, and includes the number of observations ($N_{\rm
  obs}$), the average signal-to-noise ratio, the date range and time
span, the mean radial velocity, and the standard deviation
$\sigma_{\rm RV}$ for each star. To be conservative, we have chosen to
adopt $\sigma_{\rm RV}$ as the uncertainty for the mean
velocities. The next column gives the predicted RV within the cluster,
to be discussed in Section~\ref{sec:clusterRV}. Some of the objects
are known to have close visual companions, detected mostly by the
lunar occultation technique or by other imaging methods. For those,
the following column in the table provides the angular separation and
magnitude difference. In all cases the periods of these astrometric
companions are likely to be decades long under reasonable assumptions
for the masses, and are generally not expected to induce measurable
velocity variations over the duration of our observations.  As seen in
the table, more than half of the stars have been claimed to be
velocity variables at one time or another over the last century, often
by more than one author, and tentative spectroscopic orbital solutions
have been published for four of them. However, the velocities on which
those assessments are based are typically of relatively poor quality
compared to those in this paper, and as it turns out, except for one
or possibly two objects, our observations do not appear to support any
of those claims of variability, or any of the published orbits. In one
case (HD~22637) we do find that the star has variable velocity, but
the orbit is completely different from the one published. In another
(HD~23338), we present a new orbit based on our own measurements
combined with those of others that is also different from the orbit
proposed in the literature, although we still view the solution as
preliminary. There is also one case (HD~24899) previously unrecognized
as a spectroscopic binary, for which we present a new orbit. In the
following we present notes with additional information for stars in
the table with conflicting claims of RV variability from different
authors, claims of double lines, previously claimed or new
spectroscopic orbits, or other features of interest.

\subsection{Notes on individual objects}

\noindent{\bf HD~22614:} The velocity variability claim by
\cite{Smith:1944} is based on only two measurements separated by 83
days, showing a difference of 40~\kms. Neither our 9 observations over
590~days nor the 3 of \cite{Pearce:1975} over 92~days display anywhere
near that level of variability.

\begin{figure}
\epsscale{1.15}
\plotone{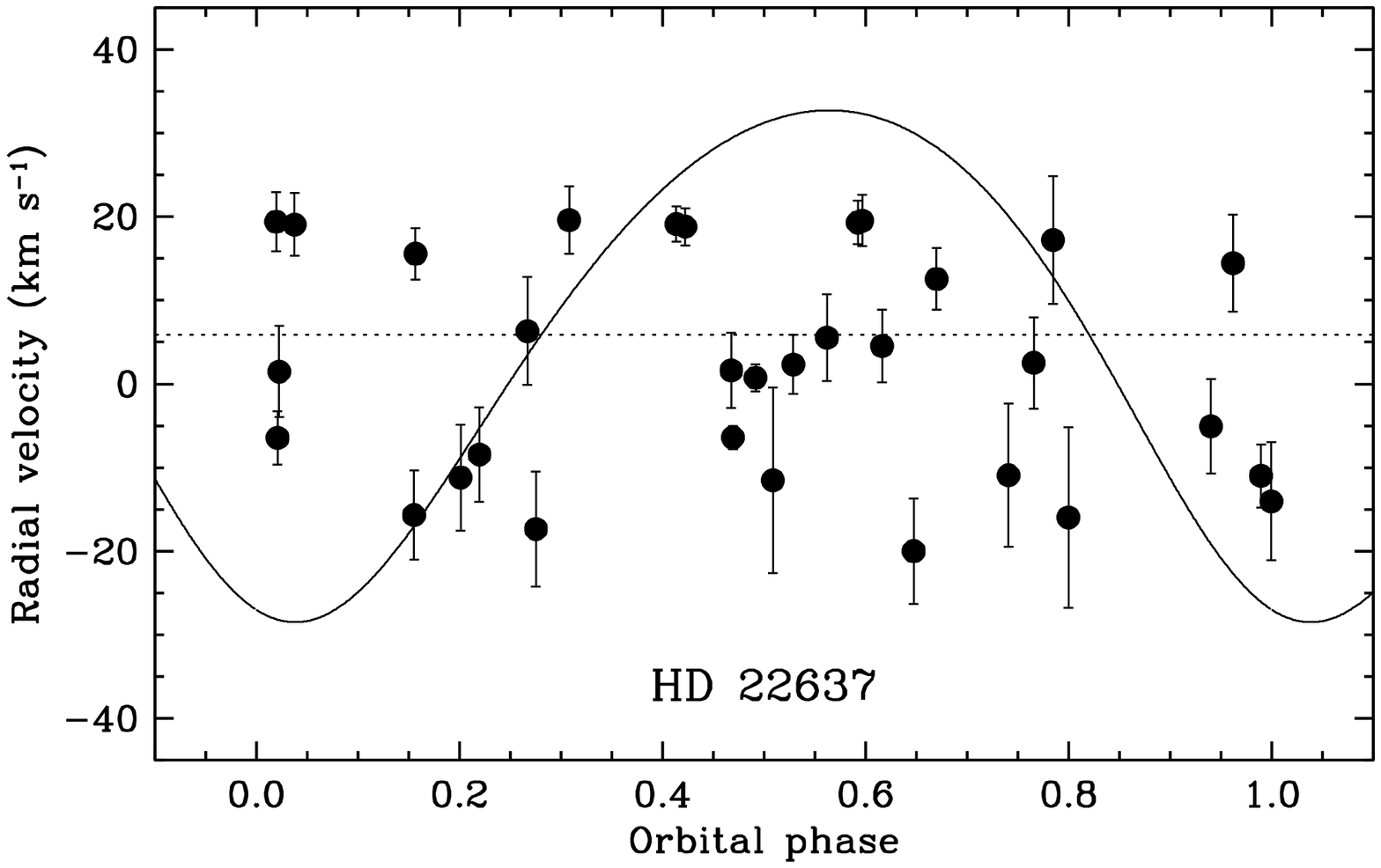}
\vskip 5pt
\plotone{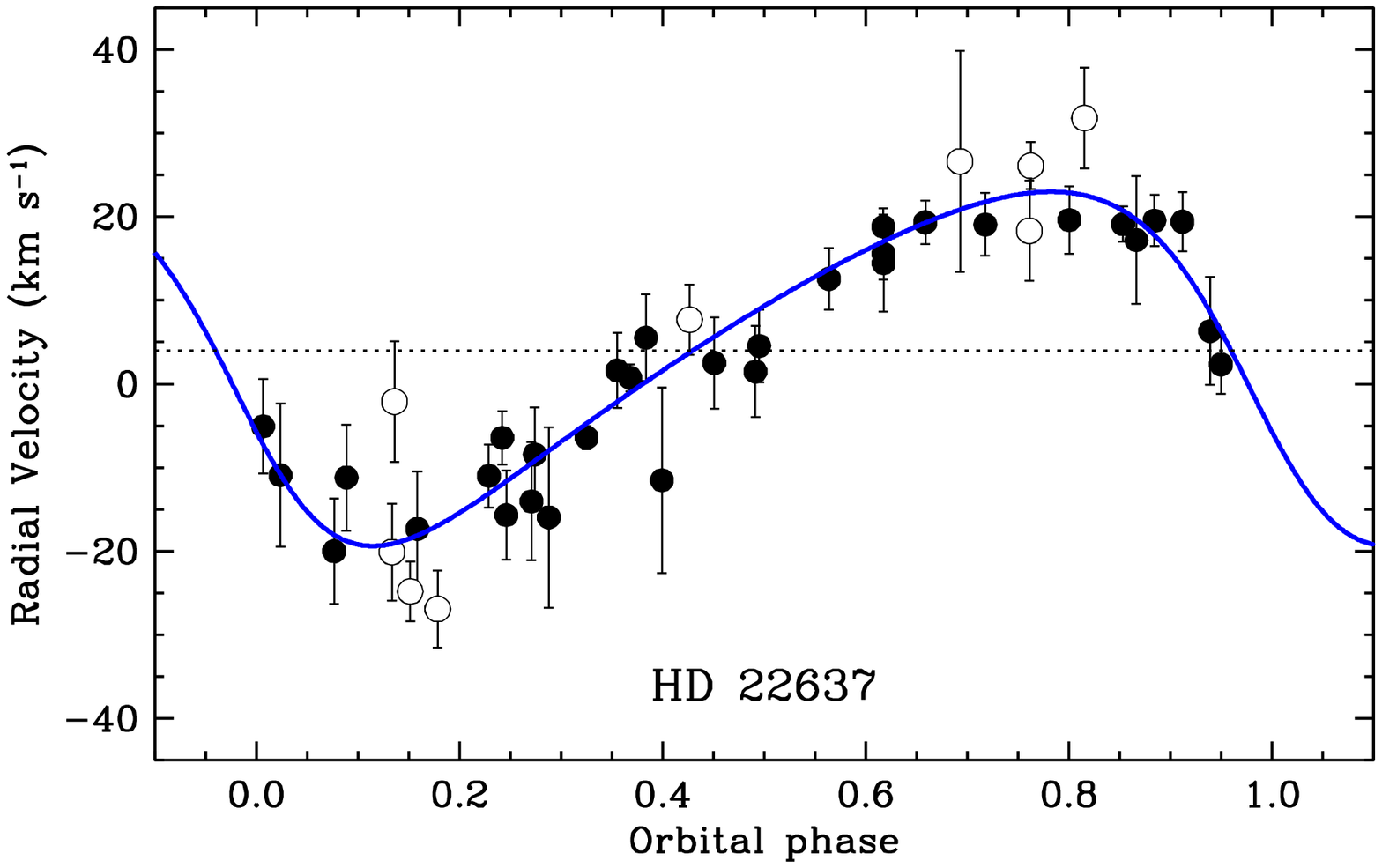}

\figcaption{\emph{Top:} Orbit model proposed by \cite{Pearce:1975} for
  HD~22637, with a period of 4.674~days, compared with our own
  measurements. The dotted line is the reported center-of-mass
  velocity of the system. Our velocities are seen not to follow
  predictions. \emph{Bottom:} New orbital solution with $P =
  71.8$~days, shown with our own measurements (filled circles) as well
  as historical velocities from the literature. Those of
  \cite{Smith:1944} have been adjusted to place them on the same
  system as the ones of \cite{Pearce:1975}, following the prescription
  from those authors. As before, the dotted line marks the
  center-of-mass velocity, and phase 0.0 corresponds to periastron
  passage. \label{fig:orbit_HD22637}}

\end{figure}

\noindent{\bf HD~22637:} \cite{Pearce:1975} reported an orbital
solution with a period of 4.674 days. Our observations are
inconsistent with this orbit (see Figure~\ref{fig:orbit_HD22637}, top
panel), and instead show clear RV variability with a much longer
period of 71.8 days. As it turns out, most of the velocities from
\cite{Pearce:1975} fit our orbit quite well, as do the older ones of
\cite{Smith:1944}. Our new spectroscopic orbit using the new and old
velocities is shown in the bottom panel of the figure, and the
elements from a weighted least-squares solution are listed in
Table~\ref{tab:orbits}. The fairly large coefficient for the minimum
secondary mass suggests that, for certain orbital inclination angles,
the companion may be massive enough and therefore bright enough to
make its lines visible in our spectra. If we adopt a primary mass of
$\sim$2.4~$M_{\sun}$ based on its spectral type, and a conservative
detection sensitivity of 1~mag for the brightness difference, we
estimate that a secondary with a mass of 1.75~$M_{\sun}$ or larger
should be bright enough to be detected in our spectra. Given the
orbital elements, this would occur for inclination angles smaller than
about 36\arcdeg, with a $\sim$19\% probability under the assumption of
random inclination angles. Our lack of detection of lines of the
secondary (see Figure~\ref{fig:coadd}) would then imply an inclination
angle larger than this limit.  The center-of-mass velocity of the
binary is lower than the cluster mean of about
6~\kms\ \citep{Liu:1991, Morse:1991, Rosvick:1992, Raboud:1998,
  Gao:2019, Gaia:2018b}, which may be a sign of another more distant
star in the system. To our knowledge no such companions have been
detected so far.  A speckle interferometry observation by
\cite{Mason:2009} yielded a negative result. At the distance to the
Pleiades, the 71.8-day orbit should subtend an angle of about 4 mas,
which is resolvable with long-baseline interferometers.

\setlength{\tabcolsep}{2pt}  
\begin{deluxetable}{lcc}
\tablecaption{Orbital Elements for Two New Spectroscopic Binaries in the Pleiades \label{tab:orbits}}
\tablehead{
\colhead{Parameter} &
\colhead{HD 22637} &
\colhead{HD 24899}
}
\startdata
$P$ (day)                                     & $71.8198 \pm 0.0084$ & $3635 \pm 19$ \\
$\gamma$ (\kms)                               & $3.94 \pm 0.78$      & $7.183 \pm 0.068$      \\
$K_1$ (\kms)                                  & $21.2 \pm 1.3$       & $3.72 \pm 0.16$       \\
$e$                                           & $0.284 \pm 0.048$    & $0.528 \pm 0.035$    \\
$\omega_1$ (degree)                           & $111 \pm 11$         & $207.8 \pm 3.4$         \\
$T_{\rm peri}$ (HJD$-$2,400,000)              & $58403.4 \pm 2.0$    & $49190 \pm 47$    \\
$M_2 \sin i / (M_1+M_2)^{2/3}$ ($M_{\sun}$)   & $0.396 \pm 0.022$    & $0.2280 \pm 0.0088$    \\
$a_1 \sin i$ ($10^6$ km)                      & $20.1 \pm 1.1$       & $157.8 \pm 6.0$       \\
$N_{\rm obs}$                                 & 32 + 9               & 16 + 5               \\
rms (\kms)                                    & 5.83                 & 0.35               
\enddata

\tablecomments{The symbols $\omega_1$ and $a_1 \sin i$ represent the
  longitude of periastron and projected semimajor axis of the primary
  component, respectively. Other symbols have their usual meaning.}

\end{deluxetable}

\noindent{\bf HD~22702:} Our RVs seem to be the first to be obtained
for this star. Our 10 measurements show no meaningful change over a
period of 2~yr.

\noindent{\bf HD~23155:} The claim by \cite{Smith:1944} that the
velocity is variable is based on only two measurements separated by 4
days, showing a change of 30~\kms. Our own 10 observations show no
significant change over nearly 590 days, and the 4 measurements of
\cite{Pearce:1975} spanning 40 days also show little variation.

\begin{figure}
\epsscale{1.18}
\plotone{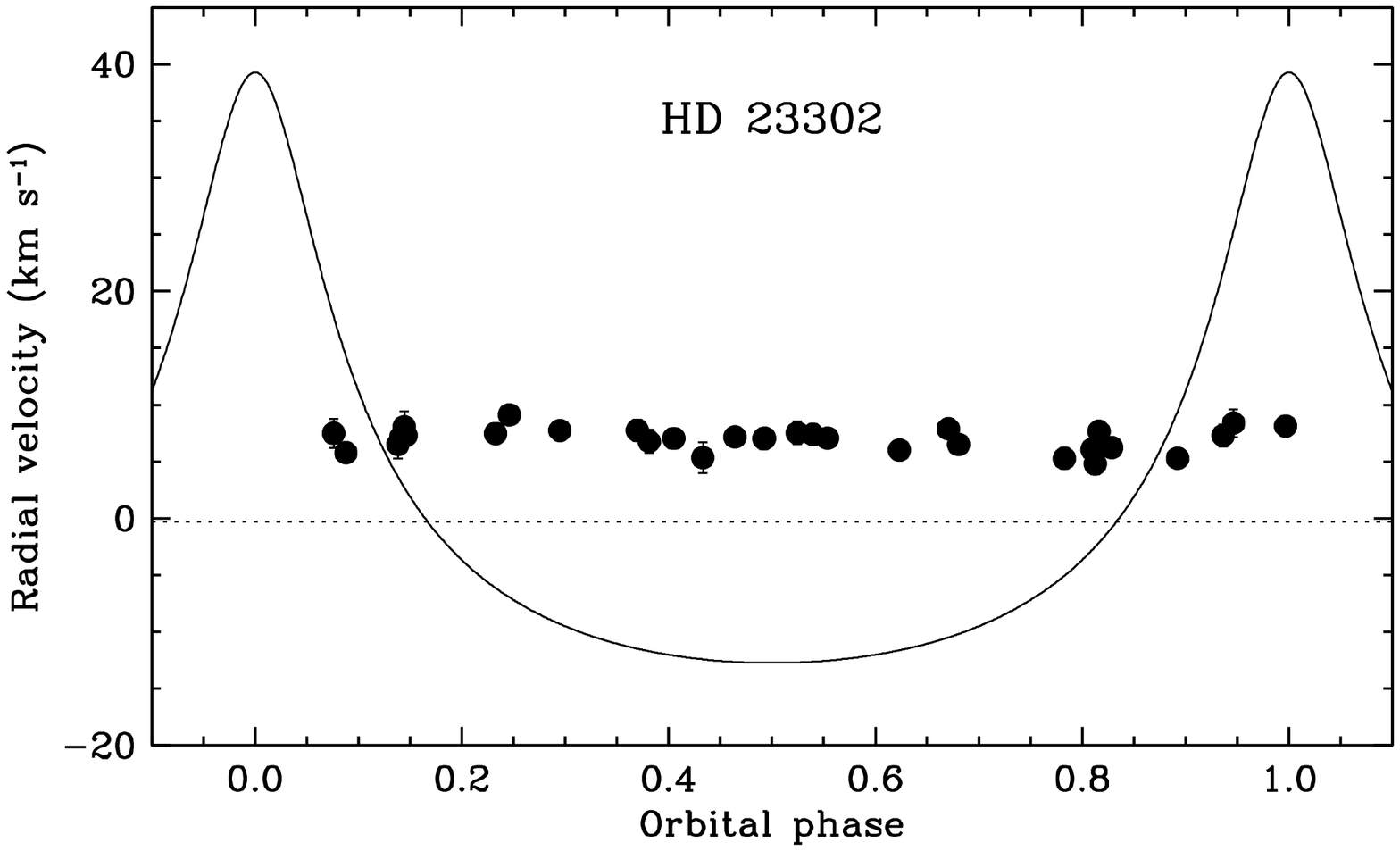}
\vskip 2pt
\plotone{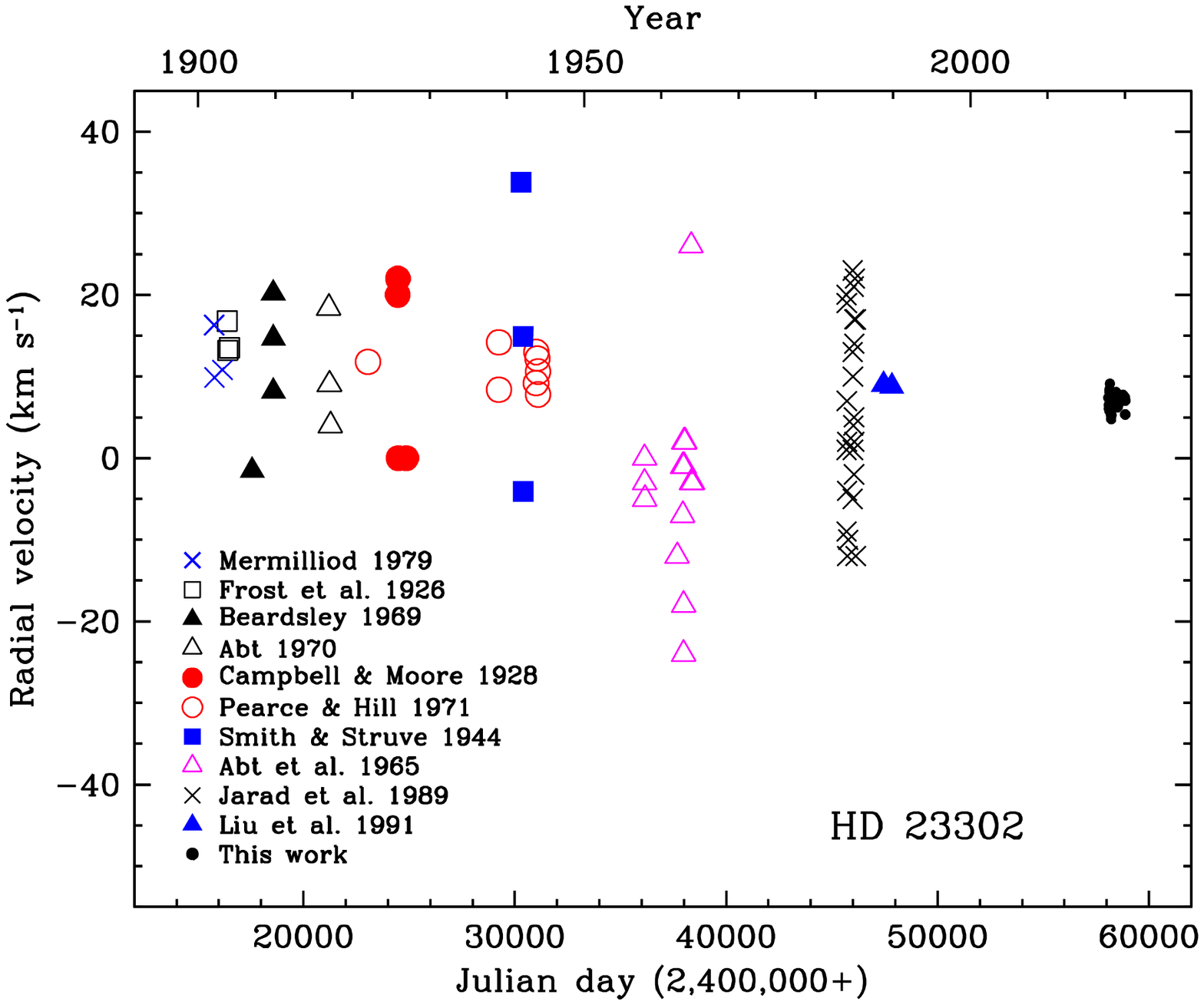}

\figcaption{\emph{Top:} Our velocity measurements for HD~23302 plotted
  against predictions from the 100.46-day orbit of \cite{Abt:1965}.
  The orbit is spurious.  \emph{Bottom:} Historical velocities for
  HD~23302 showing a large scatter typical of B and A stars in the
  Pleiades, along with our own measurements. 
\label{fig:HD23302}}

\end{figure}

\noindent{\bf HD~23302:} This is Electra (17~Tau). \cite{Abt:1965}
reported a tentative 100.46 day spectroscopic orbit with a velocity
semiamplitude of 26\,\kms, based on their own observations and others
from the literature, but noted that it was poorly determined.
\cite{Pearce:1971} were skeptical, as their independent observations
did not follow that orbit particularly well. Our own, more precise
observations show no significant velocity variability over a nearly
800-day time span, ruling out the orbit. \cite{Chini:2012} also found
the star to be constant in RV from 5 unpublished measurements. We
demonstrate the conflict between our observations and the Abt orbit in
the top panel of Figure~\ref{fig:HD23302}. \cite{Jarad:1989} also
doubted the orbit based on their new RV measurements, but proposed two
alternative solutions with much shorter periods of 3.83 and 4.29~days,
and semiamplitudes of about 13 and 10~\kms, respectively. Although
full details were not given, our constant velocities clearly reject
both of those orbits as well. These claims of RV variability in
rapidly rotating stars are not surprising.  Even for a bright and
easily observable object such as HD~23302 ($V = 3.70$), the very large
dispersion of the historical velocities, compounded by differences in
velocity zero-points between the various data sets, makes it tempting
to interpret the scatter as implying real variability if the
observational uncertainties are not well known, or are underestimated.
The lower panel of Figure~\ref{fig:HD23302} gathers all available
velocities for HD~23302, and is representative of the situation common
to many of the other stars in our sample.  Corresponding figures for
other objects that have also been claimed or suspected to have
variable RVs look qualitatively the same, and are not shown.  Finally,
we note that a visual companion to HD~23302 is known from lunar
occultation observations \citep[e.g.,][]{Richichi:1996}, at a
projected separation of about 0\farcs2 and with a brightness
difference of $\Delta K = 3.5$. The orbital period is expected to be
of the order of a century.

\noindent{\bf HD~23324:} 18~Tau. \cite{Abt:1965} suspected double
lines at two of their epochs, but reported the velocities were not
measurable. \cite{Pearce:1975} also mentioned double lines. We see no
evidence of this in our spectra (see Figure~\ref{fig:coadd}). We are
also unable to find any significant periodicity in the historical
velocities.  Our own observations show no change over nearly 600 days,
and neither do 9 unpublished RVs from \cite{Chini:2012}. A close
companion is known from lunar occultations. It may be responsible for
the fact that our mean velocity is about 2~\kms\ higher than the
cluster mean.

\begin{figure}
\epsscale{1.15}
\plotone{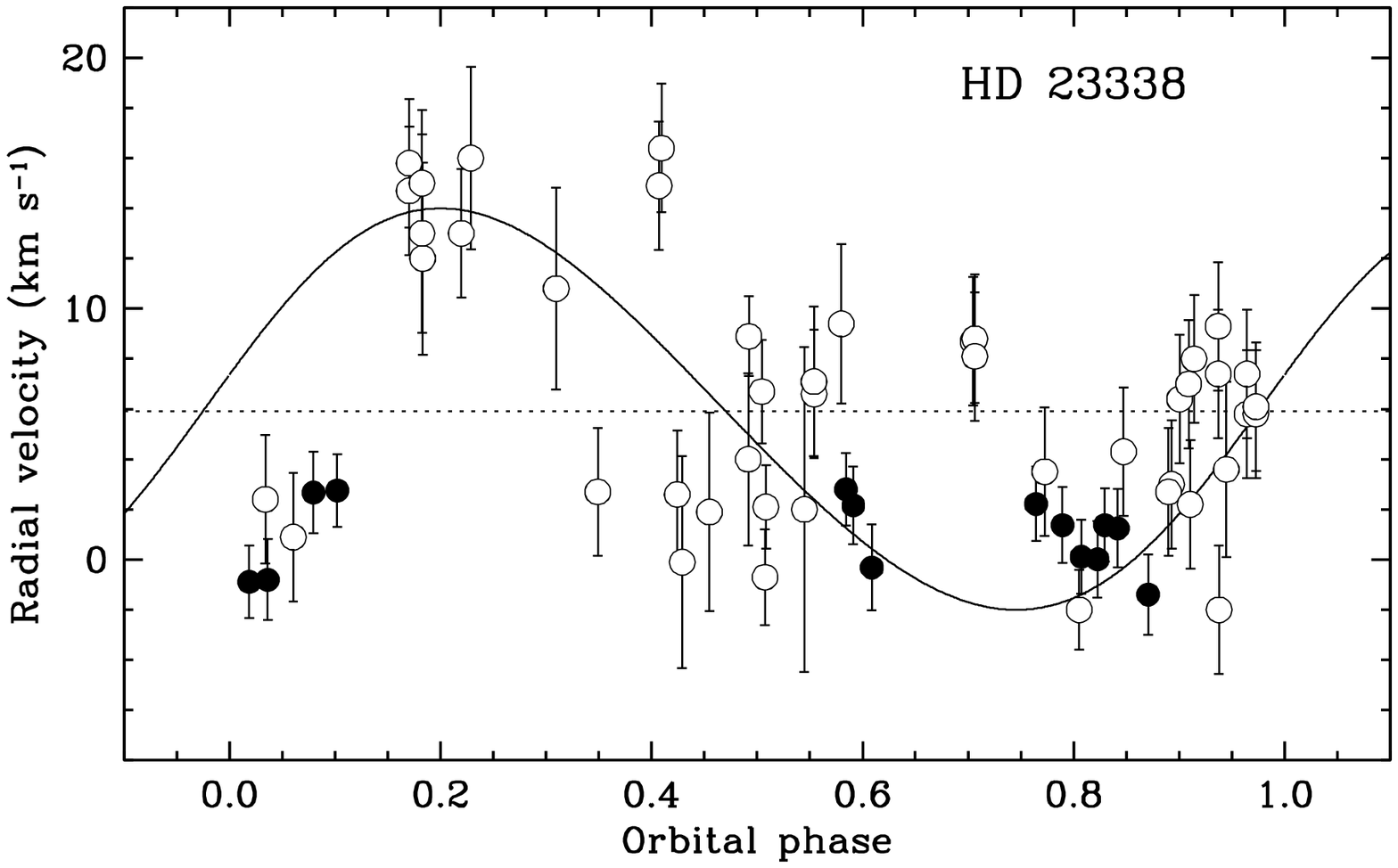}
\vskip 5pt
\plotone{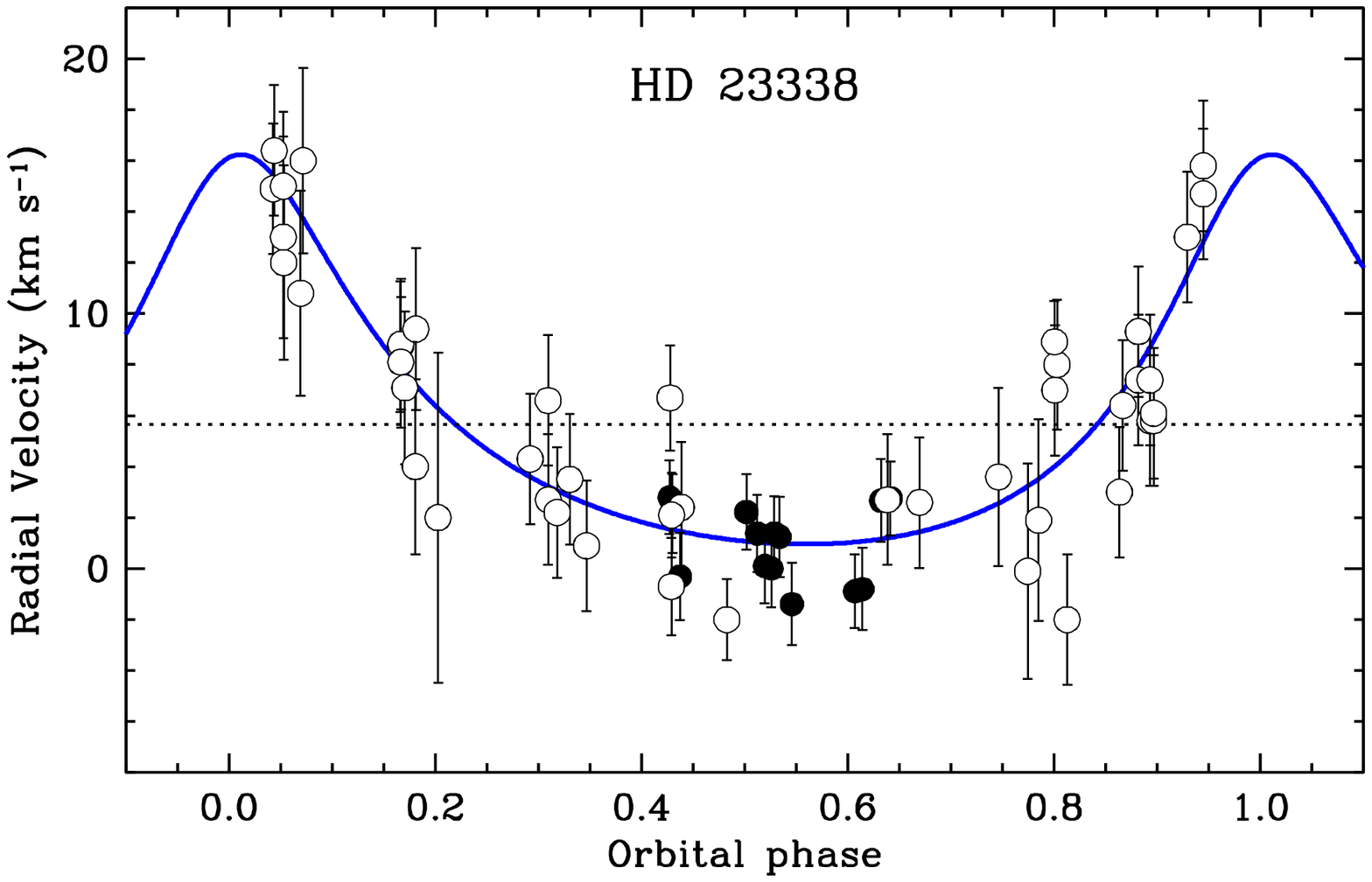}

\figcaption{\emph{Top:} Orbit model for HD~23338 proposed by
  \cite{Abt:1965}, with a period of 1313~days, compared with our own
  measurements (shown as filled symbols) and those of others
  \citep[][all represented with open symbols]{Adams:1904, Jung:1914,
    Campbell:1928, Smith:1944, Abt:1965, Abt:1970, Pearce:1971,
    Pearce:1975, Andersen:1983, Liu:1991}. The observations are seen
  not to follow predictions. The dotted line is the proposed
  center-of-mass velocity of the system. \emph{Bottom:} New orbital
  solution (preliminary) with $P = 3172.4$~days based on the same
  velocities as above. The measurements of \cite{Smith:1944} have been
  adjusted to place them on the same system as the ones of
  \cite{Pearce:1975}, following the prescription from those authors.
  As before, the dotted line marks the center-of-mass velocity, and
  phase 0.0 corresponds to periastron
  passage. \label{fig:orbit_HD23338}}

\end{figure}

\noindent{\bf HD~23338:} This is Taygeta, or q~Tau. \cite{Abt:1965}
reported a tentative orbit with $P = 1313$ days and a velocity
semiamplitude of 8\,\kms, but pointed out that it was not completely
convincing. \cite{Pearce:1971} cast doubt on the binary nature of the
object, as their own observations were essentially constant.
\cite{Chini:2012} reached a similar conclusion based on 8 spectra.
Our own velocities over an interval of 680 days also show little
change (see Figure~\ref{fig:orbit_HD23338}, top panel), but we note
that they are all significantly lower than the mean velocity for the
cluster, averaging $0.95 \pm 1.48$~\kms.  HD~23338 is known from lunar
occultation measurements to have a companion about 2.3~mag fainter in
the $K$ band.  \cite{Richichi:1994} discussed the available
measurements, some of which they considered of marginal quality, and
from two events they viewed as the most reliable they inferred a true
angular separation of about 0\farcs19 at a mean epoch of 1989.29, and
a position angle of 156\arcdeg. At the distance of the Pleiades, this
separation corresponds to an orbital period of decades, which could
well explain both why our velocities seem constant and why their
average differs by $\sim$5\,\kms\ from the cluster mean.

We have re-examined the historical velocities for HD~23338 going back
more than a century, and discovered that it is possible to obtain a
spectroscopic orbital solution with a period of about 8.7~yr that
satisfies most of the available RVs, including our own. This solution
is shown graphically in the bottom panel of
Figure~\ref{fig:orbit_HD23338}, and the elements are listed in
Table~\ref{tab:elem_HD23338}. The RV observations span nearly 118~yr,
and cover 13.6 cycles of the binary. At the present time we consider
this solution for HD~23338 to be preliminary, for two reasons. First,
a few measurements by several authors were found to deviate
significantly from the fit, and were excluded: one by
\cite{Jung:1914}, one by \cite{Smith:1944}, 5 by \cite{Abt:1965}, and
also all 4 measurements by \cite{Beardsley:1969} (one being very
uncertain). Velocities from the latter source tend to show more
scatter than those by other authors.  Second, the solution was found
to be somewhat sensitive to the weighting of the observations. Many of
the old measurements have no published errors, and we have arbitrarily
assigned them uncertainties of 3~\kms. For the ones that do, we have
converted from the published probable errors to mean errors where
necessary. And to guard against the possibility that our own errors
are underestimated, we also added 2~\kms\ in quadrature to all
observations. While changes to this error scheme do not affect the
orbital period very much, they do impact the eccentricity and velocity
semiamplitude. Confirmation of our tentative solution would therefore
benefit from new spectroscopic observations during the next periastron
passage, which we predict should occur in early 2023.

\setlength{\tabcolsep}{6pt}  
\begin{deluxetable}{lc}
\tablecaption{Preliminary Spectroscopic Orbital Elements for HD 23338 \label{tab:elem_HD23338}}
\tablehead{
\colhead{Parameter} &
\colhead{value}
}
\startdata
$P$ (day)                                     & $3172.4 \pm 9.9$ \\
$\gamma$ (\kms)                               & $5.66 \pm 0.34$      \\
$K_1$ (\kms)                                  & $7.64 \pm 0.88$       \\
$e$                                           & $0.391 \pm 0.079$    \\
$\omega_1$ (degree)                           & $350 \pm 11$         \\
$T_{\rm peri}$ (HJD$-$2,400,000)              & $37784 \pm 82$    \\
$M_2 \sin i / (M_1+M_2)^{2/3}$ ($M_{\sun}$)   & $0.485 \pm 0.044$    \\
$a_1 \sin i$ ($10^6$ km)                      & $307 \pm 28$       \\
$N_{\rm obs}$                                 & 14 + 45              \\
rms (\kms)                                    & 2.44               
\enddata

\tablecomments{The symbols $\omega_1$ and $a_1 \sin i$ represent the
  longitude of periastron and projected semimajor axis of the primary
  component, respectively. Other symbols have their usual meaning. A
  total of 45 observations from the literature have been included in
  this orbital solution, along with 14 of our own.}

\end{deluxetable}

Adopting a mass for the primary of 4~$M_{\sun}$ based on its spectral
type, and a range of secondary masses, we estimate the angular
semimajor axis of this preliminary orbit to be about 0\farcs06. The
predicted angular separation at the epoch of the first of the lunar
occultation measurements discussed by \cite{Richichi:1994} (1988.97)
is then $\sim$0\farcs04, with little dependence on the unknown
inclination angle. This is smaller than the $0\farcs0777 \pm
0\farcs0005$ projected separation recorded during that
event\footnote{Lunar occultation events do not yield the true angular
  separation of a binary; only its projection along the direction of
  the lunar motion. The true separation at this epoch could therefore
  be larger than 0\farcs0777, making it even more inconsistent with
  the prediction from the spectroscopic orbit.}. Thus, the
spectroscopic companion does not appear to be the same as the
astrometric companion.  We point out also that, in order to infer a
true angular separation of 0\farcs19 from two one-dimensional lunar
occultation measurements at epochs 1988.97 and 1989.72,
\cite{Richichi:1994} had to assume the companion did not move
appreciably in the intervening 9 months.  However, to the extent that
our spectroscopic orbit is correct, it indicates that the companion
did in fact move significantly over that period, which may invalidate
the 0\farcs19 estimate of the true angular separation. Additional
astrometric observations resolving the wide pair are needed to gain a
better understanding of its orbit.

\noindent{\bf HD~23323:} We are not aware of any RV measurements for
this star in the literature other than our own. We see no significant
change over more than 4~yr.

\noindent{\bf HD~23361:} No significant periodicity could be found in
the historical velocities for this star. Our 10 measurements over more
than 5~yr do not support the variability claims in the literature.

\begin{figure}
\epsscale{1.15}
\plotone{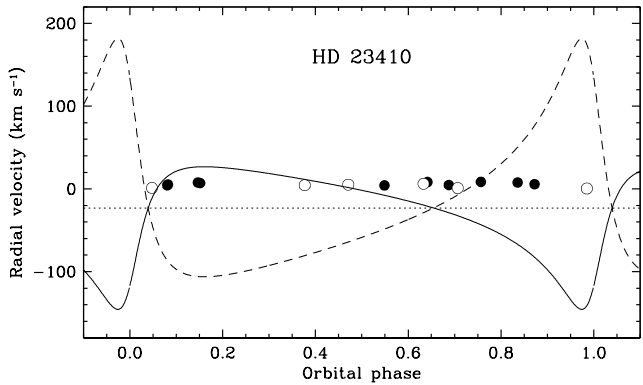}

\figcaption{Double-lined spectroscopic orbit model for HD~23410
  proposed by \cite{Abt:1965}, with a period of 7.15~days, compared
  against our own measurements (shown as filled symbols) and those of
  \cite{Smith:1944} and \cite{Pearce:1975} (open symbols). The
  observations are seen not to follow predictions. The dotted line
  represents the proposed center-of-mass velocity of the
  system. \label{fig:orbit_HD23410}}

\end{figure}

\noindent{\bf HD~23410:} A double-lined orbit for this object was
reported by \cite{Abt:1965}, with a period of 7.15 days, a large
eccentricity, and large semiamplitudes of 86 and 144\,\kms\ for the
primary and secondary (with the latter one being more uncertain). The
authors indicated it was based only on the \ion{Ca}{2} K line, and
that the components are of nearly the same brightness, leading to
confusion and uncertainty in deriving the orbital elements. The
center-of-mass velocity they gave, $-23$\,\kms, is inconsistent with
its well established cluster membership \citep{Gao:2019}, and indeed
our velocities show no change from an average of 6.4\,\kms\ over about
a year and a half, indicating the proposed orbit is spurious.  Double
lines are not apparent in our spectra (see Figure~\ref{fig:coadd}).
Measurements by \cite{Smith:1944} and by \cite{Pearce:1975} also
disagree with the orbit. We illustrate this in
Figure~\ref{fig:orbit_HD23410}. This is a known visual binary with a
separation of about 3\farcs5, and a brightness difference of about
3~mag in the visible.

\noindent{\bf HD~23409:} \cite{Liu:1991} suspected double lines, but
this is not obvious in our spectra (see Figure~\ref{fig:coadd}). Our
RV measurements show no significant change over nearly 6 years.  We
are unable to find any coherent periodicity in the previously
published velocities.

\noindent{\bf HD~23489:} \cite{Abt:1965} reported that although the
scatter appeared large, they could not discover any periodicity in the
velocities.  Seven other measurements have been published since, but
no orbit has emerged. Our own velocities are constant. A wide
companion is known from speckle interferometry.

\noindent{\bf HD~23512:} Although \cite{Smith:1944} thought the
velocity might be variable, \cite{Abt:1965} saw no change in their own
measurements.  \cite{Liu:1991} suspected double lines; we see no
evidence of that (see Figure~\ref{fig:coadd}). Our own velocities are
constant within the measurement uncertainties, over nearly 2
years. The star is listed as a visual binary with a separation of
about 0\farcs1 and a brightness difference of 2~mag
\citep{Mason:2001}.

\noindent{\bf HD~23585:} \cite{Liu:1991} reported a 15~\kms\ RV change
from two observations separated by 700 days. We see no change in our
24 measurements over 9 yr, nor are we able to detect any periodicity
in the historical velocities.

\noindent{\bf HD~23629:} 24~Tau. \cite{Abt:1965} cast doubt on the
velocity variability claim of \cite{Smith:1944}. We also see little
change in the RVs over a year and a half. The detection of a lunar
occultation companion 1.1~mag fainter in $B$ was reported by
\citep{McGraw:1974}, at a projected separation of 0\farcs0019.

\noindent{\bf HD~23643:} Only \cite{Liu:1991} have claimed this is a
velocity variable, based on two measurements about 700~days apart
showing a difference of 23~\kms. Neither our observations nor those of
others confirm this.

\noindent{\bf HD~23753:} While some authors have considered this star
to be a velocity variable, \cite{Chini:2012} reported a constant
velocity from their 6 spectra. Our own measurements over about a year
also show little change. A companion is known from lunar occultations.
The rotational velocity of this star is the largest in our sample.

\noindent{\bf HD~23763:} Although RV variability has been claimed for
this object, we are unable to find a satisfactory orbital solution
using the previously published velocities. Our own 23 measurements
over 9 years show no change.

\noindent{\bf HD~23863:} Another object with no significant RV
variation from our own observations over nearly 8 yr, and a large
scatter but no coherent motion that we can discern in the historical
RVs.

\noindent{\bf HD~23950:} \cite{Hube:1970} flagged the star as a
velocity variable, but \cite{Chini:2012} reported their 7 measurements
show no significant change.  We also see no RV variation in almost
8~yr. Our mean velocity is higher than the cluster mean. The changes
seen by \cite{Hube:1970} are likely due to measurement errors, as two
of their published velocities obtained on the same night within half
an hour of each other differ by 23~\kms.

\noindent{\bf HD~24013:} \cite{Liu:1991} claimed the detection of
double lines. This is a known non-member, which {\it Gaia\/} confirms
from a parallax and proper motion that disagree with the mean values
for the cluster. It was observed here only because it was on the
original list of possible cluster members with which the project
started.  Our velocities show little change within the errors over 660
days, and we see no sign of double lines in our spectra
(Figure~\ref{fig:coadd}).  \cite{Qian:1991} reported the detection of
a companion from lunar occultations that is 1.1~mag fainter than the
primary in $B$, at a projected separation of 0\farcs0155.

\noindent{\bf HD~24711:} Aside from our own 5 measurements, we are
aware of only 3 others in the literature: two from \cite{Smith:1944}
and one from \cite{Pearce:1975}. The two from \cite{Smith:1944} are 81
days apart and differ by 33~\kms. Ours show little scatter over 324
days.

\begin{figure}
\epsscale{1.15}
\plotone{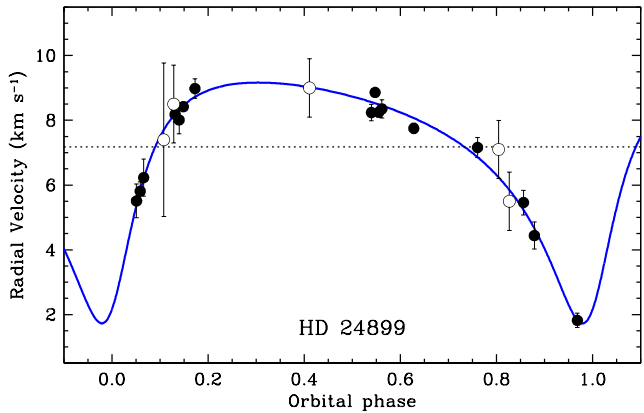}

\figcaption{New orbital solution for HD~24899. Our own measurements
  are shown with filled symbols, and those of \cite{Smith:1944} and
  \cite{Pearce:1975} with open symbols. The \cite{Smith:1944} have
  been adjusted to place them on the same system as the ones of
  \cite{Pearce:1975}, following the prescription from those authors.
  The dotted line marks the center-of-mass velocity, and phase 0.0
  corresponds to periastron passage. \label{fig:orbit_HD24899}}

\end{figure}

\noindent{\bf HD~24899:} There is a clear long-term variability in the
velocities from this work, with a period of about 10 yr. This appears
to have gone unnoticed until now. Three velocity measurements from
\cite{Pearce:1975} and two from \cite{Smith:1944} are actually
consistent with the trend.  The new orbital solution incorporating all
measurements is shown in Figure~\ref{fig:orbit_HD24899}, and the
elements are listed in Table~\ref{tab:orbits}. If we assume that the
primary has a mass near 2.6~$M_{\sun}$ based on its spectral type, and
if we adopt the same conservative sensitivity limit to companions as
used for HD~22637 (1~mag), we estimate that the minimum mass for the
secondary to be bright enough to be detected is about 1.9~$M_{\sun}$.
For this to be the case the inclination angle must be smaller than
about 19\arcdeg, which would be expected to occur only $\sim$5\% of
the time for an isotropic distribution. This seems consistent with the
fact that we see no evidence of double lines (Figure~\ref{fig:coadd}).
The {\it Hipparcos\/} mission detected and measured a significant
astrometric acceleration (i.e., curvature in the proper motion) over
the 2~yr duration of the observations for this star, which is likely a
reflection of the orbit we report here. The same effect was detected
from proper motion differences between {\it Hipparcos\/} and {\it
  Tycho-2\/} by \cite{Makarov:2005}, and between {\it Hipparcos\/} and
{\it Gaia\/} by \cite{Kervella:2019}. To our knowledge there has been
no direct detection of the companion as yet.  At the distance to the
Pleiades, we expect the angular separation to be of the order of
50~mas.

\subsection{The mean velocity of the cluster}
\label{sec:clusterRV}

\begin{figure}
\epsscale{1.15}
\plotone{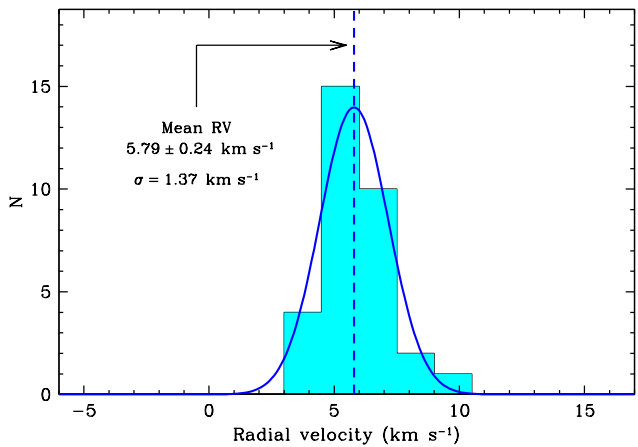}

\figcaption{Histogram and Gaussian representation of the 32 mean
  radial velocities for the B, A, and F stars in our Pleiades sample.
  HD~24013 is not a member of the cluster, and has been
  excluded.\label{fig:meanrvs}}

\end{figure}

The radial velocities from Table~\ref{tab:rvs} for the B, A, and F
stars in our sample fall close to the known mean value for the
Pleiades of about 6~\kms\ (see Figure~\ref{fig:meanrvs}). Ignoring
binarity for the moment, the average of the 32 stars, with the
non-member HD~24013 excluded, is $5.79 \pm 0.24~\kms$. This is in very
good agreement with determinations by others \citep[e.g.,][]{Gao:2019,
  Gaia:2018b}. However, the standard deviation of 1.37~\kms\ is much
larger than the internal velocity dispersion of the cluster, which is
estimated to be only about 0.5~\kms\ \citep[e.g.,][]{Jones:1970,
  Rosvick:1992, Makarov:2001}.

\begin{figure}
\epsscale{1.15}
\plotone{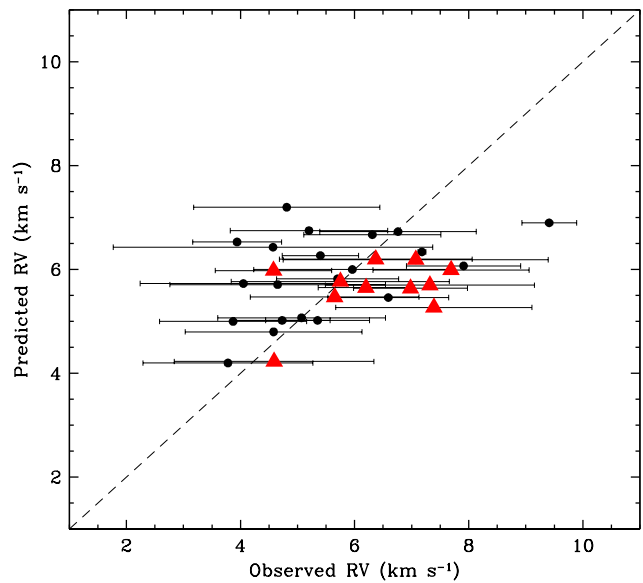}

\figcaption{Measured radial velocities for each of our stars compared
  with the predicted velocities within the cluster based on the
  convergent-point solution from the {\it Gaia\/} mission
  \citep{Gaia:2018b}. The dashed line represents the one-to-one
  relation. Red triangles mark stars that have known astrometric
  companions. \label{fig:clusterRV}}

\end{figure}

There are at least three possible reasons for the larger scatter. One
is of course measurement errors, which we address below. Another is
that the angular extent of the Pleiades on the sky is quite large
($\sim$10\arcdeg), and changes in the projection of the mean space
velocity of the cluster along the line of sight from one end to the
other will cause a radial velocity gradient of several \kms. To show
this more quantitatively, we have listed the predicted radial velocity
of each star in Table~\ref{tab:rvs}, calculated from the position of
the convergent point and the mean distance and proper motion of the
Pleiades from the {\it Gaia\/} mission \citep{Gaia:2018b}. The
uncertainty in those predictions is negligible for our purposes ($<
0.1$~\kms).  We find, however, that even after accounting for this
gradient, the RV dispersion for our stars is 1.52~\kms, not very
different than before.  Figure~\ref{fig:clusterRV} shows the predicted
RV of each star against the observed velocities. With few exceptions
the agreement is very good. Objects with known astrometric companions
are represented with triangles.

A third possible reason for the large scatter around the cluster's
mean velocity is binarity, whether recognized or not. For the two new
confirmed spectroscopic binaries (HD~22637 and HD~24899), the
velocities we are using correspond to center-of-mass, so the effect of
the companions is already accounted for. In addition, there are 11
other targets known to have close astrometric companions, as indicated
in Table~\ref{tab:rvs}. This excludes HD~24013, and we do not count
HD~24899 either because the astrometric signature mentioned in the
notes corresponds to the same companion detected spectroscopically. In
principle it is possible that these 11 wide companions could be
affecting the measured velocities to some extent, although the
evidence for this is not always obvious in the table.  In any case, we
choose to remove them as a precaution.  Note that we also remove
HD~23338 even though it has a spectroscopic orbit, and even though its
center-of-mass velocity agrees with the prediction, because the
spectroscopic companion appears to be different from the astrometric
one (see previous section).  The average radial velocity of the
remaining 21 objects is $5.52 \pm 0.31~\kms$, close to the previous
value, and the dispersion is not very different than before:
1.44~\kms.  The excess over the internal dispersion for the cluster,
$\sigma_{\rm excess} \sim \sqrt{1.44^2 - 0.5^2} \approx 1.35~\kms$, is
comparable to our measurement errors, which average 1.26~\kms\ over
the 21 stars considered.  We may conclude, then, that our RV
measurements are not inconsistent with a small internal velocity
dispersion for the cluster, of the order of that determined by the
authors listed above, i.e., that the measurement errors play a
significant role and the discrepancy between the observed scatter and
the true internal velocity dispersion is only apparent.

\section{Rotation results}
\label{sec:rotresults}

\begin{figure*}
\epsscale{1.18}
\plotone{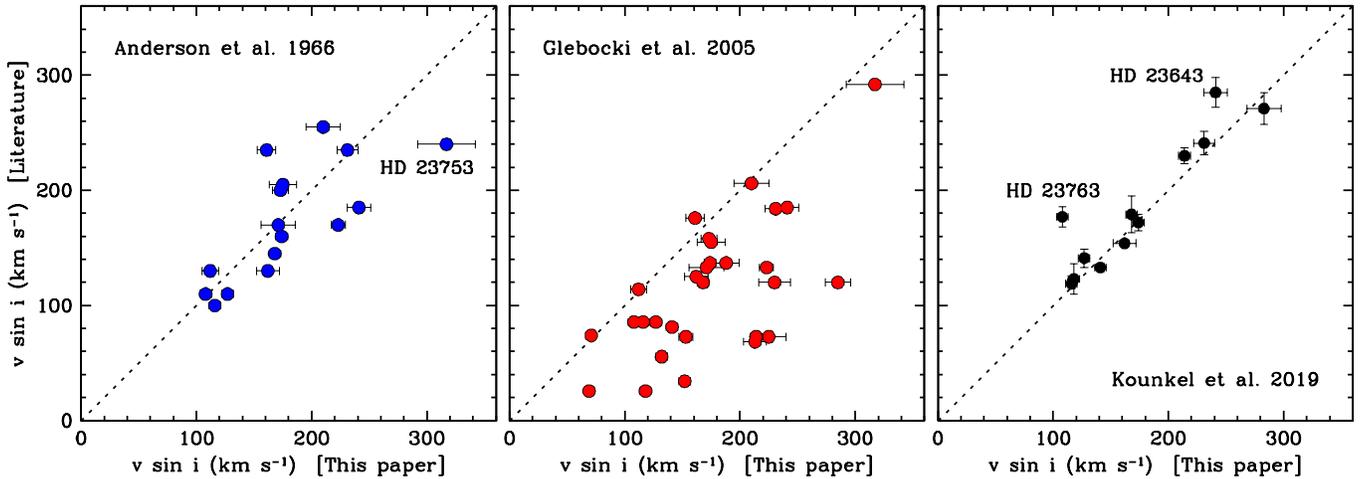}

\figcaption{Rotational velocities determined in this paper compared
  against other sources from the literature, as indicated in each
  panel. The dotted lines represent the one-to-one relations. Comments
  on a few labeled outliers are given in the text.\label{fig:vsini}}

\end{figure*}

Our $v \sin i$ determinations are presented in Table~\ref{tab:vsini},
together with other measurements from sources in the literature with
the most stars in common with our sample. Three of those sources are
compared separately in Figure~\ref{fig:vsini} against our own
measurements. Of those three, the determinations of
\cite{Anderson:1966} and the more recent ones of \cite{Kounkel:2019}
are independent, and can each be considered to have been carried out
in a homogeneous way. On the other hand, the values from the large
catalog of \cite{Glebocki:2005} are averages of historical
measurements by many different authors, including
\cite{Anderson:1966}.  Even though \cite{Glebocki:2005} made an effort
to place all of the determinations on a common system, the large
variety of measurement techniques and instrumentation involved makes
that task exceedingly difficult.  This may explain the poor agreement
with the determinations in the present paper (middle panel of
Figure~\ref{fig:vsini}).  Additionally, the VizieR database
\citep{Ochsenbein:2000} warns about possible misidentifications caused
by erroneous coordinates in some of the original sources of the
catalog.

\setlength{\tabcolsep}{5pt}  
\begin{deluxetable*}{ccccccccl}[b!]
\tablecaption{Rotational Velocity Measurements for our Sample \label{tab:vsini}}
\tablehead{
\colhead{Name} &
\colhead{$v \sin i$} &
\colhead{S1944} &
\colhead{A1965} &
\colhead{A1966} &
\colhead{A2002} &
\colhead{G2005} &
\colhead{K2019} &
\colhead{Other}
}
\startdata
 HD 21744  &  $140 \pm 5$   & \nodata & \nodata & \nodata & \nodata & \nodata & \nodata        & Ku:152   \\
 HD 22578  &  $230 \pm 14$  &   150   & \nodata & \nodata & \nodata &  120    & \nodata        &    \\
 HD 22614  &  $118 \pm 5$   &    0    & \nodata & \nodata & \nodata & 25.7    &  $123 \pm 13$  & Ku:161   \\
 HD 22637  &  $132 \pm 5$   &    50   & \nodata & \nodata & \nodata & 55.6    & \nodata        &    \\
 HD 22702  &  $148 \pm 6$   & \nodata & \nodata & \nodata & \nodata & \nodata & \nodata        & K:146(2) \\
 HD 23155  &  $214 \pm 5$   &    75   & \nodata & \nodata & \nodata & 72.8    &  $230 \pm 7$   &    \\
 HD 23302  &  $175 \pm 12$  &   200   &  230    &  205    & 85, 90  &  155    & \nodata        & F:170(12); S:180; V:215; Y:189(7); Z:170(16) \\
 HD 23324  &  $210 \pm 15$  &   150   &  235    &  255    &  185    &  206    & \nodata        & Z:212(8) \\
 HD 23338  &  $112 \pm 7$   &   150   &  140    &  130    &  105    &  114    & \nodata        & Ku:246; S:107, 108; Z:118(8) \\
 HD 23323  &  $138 \pm 5$   & \nodata & \nodata & \nodata & \nodata & \nodata & \nodata        &    \\
 HD 23361  &  $231 \pm 9$   &   100   & \nodata &  235    & \nodata &  184    &  $241 \pm 10$  & M:186.0, Mo:150  \\
 HD 23388  &  $213 \pm 10$  &    50   & \nodata & \nodata & \nodata & 68.5    & \nodata        &    \\
 HD 23402  &  $285 \pm 11$  &   150   & \nodata & \nodata & \nodata &  120    & \nodata        &    \\
 HD 23410  &  $173 \pm 7$   &   100   &  185    &  200    & \nodata &  158    & \nodata        &    \\
 HD 23409  &  $223 \pm 6$   &    75   & \nodata &  170    & \nodata &  133    & \nodata        &    \\
 HD 23432  &  $161 \pm 8$   &   150   &  210    &  235    &  160    &  176    & \nodata        & S:158, 159; Z:183(8) \\
 HD 23430  &  $141 \pm 5$   &    75   & \nodata & \nodata & \nodata & 81.3    &  $133 \pm 4$   & Ku:149   \\
 HD 23489  &  $127 \pm 5$   &    50   & \nodata &  110    & \nodata & 85.6    &  $141 \pm 8$   &    \\
 HD 23512  &  $168 \pm 5$   &    50   &  155    &  145    & \nodata &  120    &  $179 \pm 16$  & Ku:276   \\
 HD 23585  &  $116 \pm 5$   &    75   & \nodata &  100    & \nodata & 85.6    &  $119 \pm 4$   & K:113(3); M:107.5 \\
 HD 23629  &  $171 \pm 15$  &   100   &  155    &  170    & \nodata &  133    & \nodata        &    \\
 HD 23643  &  $241 \pm 10$  &   100   & \nodata &  185    & \nodata &  185    &  $285 \pm 13$  & R:219; Ro:175(9) \\
 HD 23753  &  $317 \pm 25$  &   300   &  305    &  240    &  290    &  292    & \nodata        & Z:335(8) \\
 HD 23763  &  $108 \pm 5$   &    50   &  100    &  110    & \nodata & 85.6    &  $177 \pm 9$   &    \\
 HD 23852  &  $153 \pm 6$   &    75   & \nodata & \nodata & \nodata & 72.8    & \nodata        &    \\
 HD 23863  &  $174 \pm 5$   &   100   & \nodata &  160    & \nodata &  137    &  $172 \pm 7$   &    \\
 HD 23912  &  $162 \pm 10$  &   100   & \nodata &  130    & \nodata &  125    &  $154 \pm 2$   &    \\
 HD 23913  &  $188 \pm 11$  &   150   & \nodata & \nodata & \nodata &  137    & \nodata        &    \\
 HD 23950  &   $71 \pm 4$   &    75   & \nodata & \nodata & 50, 60  & 73.9    & \nodata        &    \\
 HD 24013  &  $283 \pm 15$  & \nodata & \nodata & \nodata & \nodata & \nodata &  $271 \pm 14$  &    \\
 HD 24178  &  $225 \pm 15$  &    75   & \nodata & \nodata & \nodata & 72.8    & \nodata        &    \\
 HD 24711  &  $152 \pm 5$   &    25   & \nodata & \nodata & \nodata & 34.2    & \nodata        &    \\
 HD 24899  &   $69 \pm 4$   &    0    & \nodata & \nodata & \nodata & 25.7    & \nodata        &   
\enddata

\tablecomments{All values are in units of \kms. Column 2 reports the
  new $v \sin i$ determinations from this work. The main literature
  sources in the columns that follow are: S1944 = \cite{Smith:1944};
  A1965 = \cite{Abt:1965}; A1966 = \cite{Anderson:1966}; A2002 =
  \cite{Abt:2002}; G2005 = \cite{Glebocki:2005}; and K2019 =
  \cite{Kounkel:2019}. Codes for the literature sources in the last
  column, which have fewer measurements of the stars in our sample,
  are:
F = \cite{Fremat:2005};
K = \cite{Kahraman:2016};
Ku = \cite{Kunder:2017};
M = \cite{Margheim:2007};
Mo = \cite{Morse:1991};
R = \cite{Royer:2002};
Ro \ \cite{Rodriguez:2000};
S = \cite{Simon-Diaz:2017};
V = \cite{vanBelle:2012};
Y = \cite{Yudin:2001};
Z = \cite{Zorec:2012, Zorec:2016}.
They are followed after the colon by the $v \sin i$ measurement
(occasionally more than one), and the uncertainty in parentheses, when
available.}

\end{deluxetable*}
\setlength{\tabcolsep}{6pt}  

The $v \sin i$ measurements of \cite{Anderson:1966}, and especially
those of \cite{Kounkel:2019} (A1966 and K2019 in
Table~\ref{tab:vsini}, respectively), are much more consistent with
ours, although there are a few outliers that we have labeled. For
example, our value for HD~23753 ($317 \pm 25$~\kms) is considerably
larger than that of A1966 (240~\kms), though we note that other
measurements from the literature are all also higher than A1966, and
agree better with ours (see Table~\ref{tab:vsini}).  For HD~23763 we
measure a lower rotational velocity ($108 \pm 5$~\kms) than K2019
($177 \pm 9$~\kms), which seems to be supported by other
determinations that are also lower. And for HD~23643 we also measure a
smaller $v \sin i$ value ($241 \pm 10$~\kms) than K2019 ($285 \pm
13$~\kms), and other published values go in the same direction.

The last column of Table~\ref{tab:vsini} collects other determinations
from recent sources that have fewer stars in common with our sample.
Uncertainties are given in parentheses when available.

\section{Discussion and final remarks}
\label{sec:conclusions}

The bright, early-type stars in the Pleiades have been observed
spectroscopically since the beginning of the 20th century. Because
they tend to rotate rapidly, the RV measurements have always been
difficult, and the resulting scatter for many of these objects has led
to claims of variability, or even tentative orbital solutions in four
cases. Most of those assertions have persisted in the literature to
this day, despite the fact that several have been doubted over the
years. The velocity measurements in the present work for a sample of
33 rapidly-rotating B, A, and early F stars represent a significant
improvement in terms of precision, and are inconsistent with nearly
all claims of variability published in the past for these objects.
They additionally rule out all four tentative orbital solutions in the
literature, one from \cite{Pearce:1975}, and three from
\cite{Abt:1965}.

Among the previously claimed velocity variables, only in the cases of
HD~22637 and HD~23338 do we detect significant RV changes, but our
spectroscopic orbits are very different from the ones published. For
HD~22637, we do not confirm the 4.67~day orbit of \cite{Pearce:1975}.
Our new solution has a completely different period of 71.8~days, which
accommodates the previous observations as well. For HD~23338, the
1313~day orbit of \cite{Abt:1965} is not supported by our data either,
or by other observations in the literature. Instead, combining the
earlier observations and our own we determine a preliminary orbit with
a period of 8.7~yr. We additionally find HD~24899 to be a single-lined
binary with an orbital period of 10~yr. Velocity variability had not
been recognized before for this object.

These three are therefore new spectroscopic binaries in the
Pleiades. For HD~22637 and HD~24899, astrometric observations from the
ground could permit the determination of the dynamical masses for the
components, independently of any models. The estimated 4~mas angular
separation of HD~22637 may be resolvable with existing long-baseline
interferometers such as CHARA, and adaptive optics or speckle
interferometry observations for HD~24899 should easily resolve the
expected $\sim$50~mas separation of this pair. In the case of
HD~23338, the spectroscopic companion appears to be different from the
previously known and wider lunar occultation companion, making this a
triple system. With an estimated 60~mas semimajor axis for the inner
orbit, the new spectroscopic companion should also be resolvable with
adaptive optics or speckle interferometer techniques. At the end of
operations, the {\it Gaia} mission is expected to provide complete
astrometric orbital solutions for the photocentric motion of all three
of these objects, which should at least yield the inclination angles
of their orbits, though not the relative semimajor axes needed to
compute assumption-free masses.

Early-type stars in the field are known to display a very high
frequency of binary and higher multiplicity systems
\citep[e.g.,][]{Duchene:2013, Moe:2017}. The latter authors estimate
that $37 \pm 6$\% of stars of spectral type late-B and A are binaries
with mass ratios $q$ larger than 0.1, and another $22 \pm 7$\% are
triple or quadruple systems, for a total of about 60\%. Among mid-B
stars this fraction rises to 76\%, and it reaches close to 100\% for
the O stars. In our sample of 32 B and A-type stars that are members
of the cluster we have found two spectroscopic binaries (HD~22637 and
HD~24899), one system that is possibly triple (HD~23338), and 10 more
targets that are known to have companions from astrometric
observations. This adds up to 12 binary systems and one triple. The
implied frequency of binaries is then $12/32 \approx 37$\%, which
happens to coincide with the estimate above. Our one triple system
yields a lower occurrence rate than expected ($1/32 \approx 3$\%). We
note, however, that most of our targets only have $\sim$1--2~yr of
spectroscopic coverage, so both of these estimates are probably only
lower limits, as additional spectroscopic binaries or triples could be
present that we are not sensitive to. Also, other astrometric binary
or multiple systems may emerge from the {\it Gaia\/} mission that
would increase the frequencies even more.

A further conclusion of the study of \cite{Moe:2017} is that the
frequency of $q > 0.1$ companions to late-B and A stars per decade of
orbital period (in units of days) peaks at around $\log P \sim 4.5$
($P \sim 90$~yr, or separations of $\sim$35~au, or $\sim$0\farcs25 at
the distance to the Pleiades), and that the peak is rather broad,
extending between $\log P \approx 3.5$--5.5 (see their Figure~37).
Two of our spectroscopic systems (HD~24899 with $P = 3635$~d, and
HD~23338 with $P = 3172$~d) fall in this regime ($\log P = 3.56$ and
3.50, respectively), as do all of the targets with astrometric
companions we are aware of. The other spectroscopic binary we have
found, HD~22637 ($P = 71.8$~d), has $\log P = 1.86$, and is the only
one shortward of the peak.  Although this sample of binaries is very
small, the period distribution of the B and A stars in the Pleiades
appears to also be consistent with the results of \cite{Moe:2017}, at
least qualitatively.

\begin{acknowledgements}

The spectroscopic observations of for this work were gathered with the
help of P.\ Berlind, Z.\ Berta, M.\ Calkins, G.\ Esquerdo, D.\ Latham,
R.\ Stefanik, and S.\ Tang.  J.\ Mink is thanked for maintaining the
CfA echelle database. The anonymous referee provided numerous helpful
comments on the original manuscript. We are also grateful to George
Zhou for assistance with the LSD technique. The author acknowledges
partial support from the National Science Foundation (NSF) through
grant AST-1509375.  The research has made extensive use of the SIMBAD
and VizieR databases, operated at the CDS, Strasbourg, France, and of
NASA's Astrophysics Data System Abstract Service.  We also acknowledge
the use of the Fourth Catalog of Interferometric Measurements of
Binary Stars \citep[][and online updates]{Hartkopf:2001}, and of the
Washington Double Star Catalog, maintained at the U.S. Naval
Observatory. The work has used data from the European Space Agency
(ESA) mission {\it Gaia} (\url{https://www.cosmos.esa.int/gaia}),
processed by the {\it Gaia} Data Processing and Analysis Consortium
(DPAC,
\url{https://www.cosmos.esa.int/web/gaia/dpac/consortium}). Funding
for the DPAC has been provided by national institutions, in particular
the institutions participating in the {\it Gaia} Multilateral
Agreement.

\end{acknowledgements}



\begin{thebibliography}


\bibitem[Abt(1970)]{Abt:1970} Abt, H.~A.\ 1970, \apjs, 19, 387

\bibitem[Abt et al.(1965)]{Abt:1965} Abt, H.~A., Barnes, R.~C., Biggs,
  E.~S., et al.\ 1965, \apj, 142, 1604

\bibitem[Abt et al.(2002)]{Abt:2002} Abt, H.~A., Levato, H., \&
  Grosso, M.\ 2002, \apj, 573, 359

\bibitem[Adams(1904)]{Adams:1904} Adams, W.~S.\ 1904, \apj, 19, 338

\bibitem[Andersen \& Nordstr\"om(1983)]{Andersen:1983} Andersen, J. \&
  Nordstr\"om, B.\ 1983, \aap, 122, 23

\bibitem[Anderson et al.(1966)]{Anderson:1966} Anderson, C.~M.,
  Stoeckly, R., \& Kraft, R.~P.\ 1966, \apj, 143, 299


\bibitem[Artyukhina \& Kalinina(1970)]{Artyukhina:1970} Artyukhina,
  N.\ M., \& Kalinina, E.\ P., 1970, Tr.\ Sternberg Astron.\ Inst.,
  39, 111

\bibitem[Beardsley(1969)]{Beardsley:1969} Beardsley, W.~R.\ 1969,
  Publications of the Allegheny Observatory of the University of
  Pittsburgh, 7, 91

\bibitem[Campbell \& Moore(1928)]{Campbell:1928} Campbell, W.\ W., \&
  Moore, J.\ H. 1928, Publications of Lick Observatory, 16, 1

\bibitem[Chen et al.(2014)]{Chen:2014} Chen, Y., Girardi, L., Bressan,
  A., et al.\ 2014, \mnras, 444, 2525

\bibitem[Chini et al.(2012)]{Chini:2012} Chini, R., Hoffmeister,
  V.\ H., Nasseri, A.\ et al. 2012, \mnras, 424, 1925



\bibitem[Duch{\^e}ne \& Kraus(2013)]{Duchene:2013} Duch{\^e}ne, G., \&
  Kraus, A.\ 2013, \araa, 51, 269

\bibitem[ESA(1997)]{ESA:1997} ESA 1997, The Hipparcos and Tycho
  Catalogues, Vol.\ 1200 (Noordwijk:ESA)

\bibitem[Fr{\'e}mat et al.(2005)]{Fremat:2005} Fr{\'e}mat, Y., Zorec,
  J., Hubert, A.-M., et al.\ 2005, \aap, 440, 305

\bibitem[Frost et al.(1926)]{Frost:1926} Frost, E.~B., Barrett, S.~B.,
  \& Struve, O.\ 1926, \apj, 64, 1

\bibitem[F\H{u}r\'esz(2008)]{Furesz:2008} F\H{u}r\'esz, G. 2008, PhD
  thesis, Univ.\ Szeged, Hungary

\bibitem[Gaia Collaboration et al.(2018a)]{Gaia:2018a} Gaia
  Collaboration, Brown, A.\ G.\ A., Vallenari, A.\ et al.\ 2018a, \aap,
  616, 1

\bibitem[Gaia Collaboration et al.(2018b)]{Gaia:2018b} Gaia
  Collaboration, Babusiaux, C., van Leeuwen, F., Barstow, M.\ A.\ et
  al.\ 2018b, \aap, 616, A10

\bibitem[Gao(2019)]{Gao:2019} Gao, X.-h., 2019, \pasp, 131, 044101

\bibitem[G{\l}{\c{e}}bocki \& Gnaci{\'n}ski(2005)]{Glebocki:2005}
  G{\l}{\c{e}}bocki, R., \& Gnaci{\'n}ski, P.\ 2005, in ESA SP-560,
  13th Cambridge Workshop on Cool Stars, Stellar Systems and the Sun,
  ed.\ F. Favata, G.\ A.\ J.\ Hussain \& B. Battrick (Noordwijk,
  Netherlands: ESA), 571

\bibitem[Gray \& Corbally(1994)]{Gray:1994} Gray, R.\ O., \& Corbally,
  C.\ J. 1994, \aj, 107, 742


\bibitem[Guerrero et al.(2020)]{Guerrero:2020} Guerrero, C.~A.,
  Rosales-Ortega, F.~F., Escobedo, G., et al.\ 2020, \mnras, 495, 806

\bibitem[Haro et al.(1982)]{Haro:1982} Haro, G., Chavira, E., \&
  Gonzalez, G.\ 1982, Boletin del Instituto de Tonantzintla, 3, 3

\bibitem[Hartkopf et al.(2001)]{Hartkopf:2001} Hartkopf, W.~I.,
  McAlister, H.~A., \& Mason, B.~D.\ 2001, \aj, 122, 3480

\bibitem[Hertzsprung(1947)]{Hertzsprung:1947} Hertzsprung, E.\ 1947,
  Annalen van de Sterrewacht te Leiden, 19, A1


\bibitem[Hube(1970)]{Hube:1970} Hube, D.~P.\ 1970, \memras, 72, 233


\bibitem[Jarad et al.(1989)]{Jarad:1989} Jarad, M.~M., Hilditch,
  R.~W., \& Skillen, I.\ 1989, \mnras, 238, 1085

\bibitem[Jones(1970)]{Jones:1970} Jones, B.~F.\ 1970, \aj, 75, 563

\bibitem[Jones(1981)]{Jones:1981} Jones, B.~F.\ 1981, \aj, 86, 290

\bibitem[Jung(1914)]{Jung:1914} Jung, J.\ 1914, Astronomische
  Mitteilungen der Universitaets-Sternwarte zu Goettingen, 17, 1

\bibitem[Kahraman Ali{\c{c}}avu{\textcommabelow s} et
  al.(2016)]{Kahraman:2016} Kahraman Ali{\c{c}}avu{\textcommabelow
  s}, F., Niemczura, E., De Cat, P., et al.\ 2016, \mnras, 458, 2307

  Hirata, R., Ito, M., et al.\ 1996, \pasj, 48, 317

\bibitem[Kervella et al.(2019)]{Kervella:2019} Kervella, P., Arenou,
  F., Mignard, F., et al.\ 2019, \aap, 623, A72

\bibitem[Kochukhov et al.(2010)]{Kochukhov:2010} Kochukhov, O.,
  Makaganiuk, V., \& Piskunov, N.\ 2010, \aap, 524, A5

\bibitem[Kounkel et al.(2019)]{Kounkel:2019} Kounkel, M., Covey, K.,
  Moe, M., et al.\ 2019, \aj, 157, 196

\bibitem[Kunder et al.(2017)]{Kunder:2017} Kunder, A., Kordopatis, G.,
  Steinmetz, M., et al.\ 2017, \aj, 153, 75

\bibitem[Latham et al.(2002)]{Latham:2002} Latham, D.\ W., Stefanik,
  R.\ P., Torres, G., et al.\ 2002, \aj, 124, 1144



\bibitem[Liu et al.(1991)]{Liu:1991} Liu, T., Janes, K.~A., \& Bania,
  T.~M.\ 1991, \apj, 377, 141

\bibitem[Makarov \& Kaplan(2005)]{Makarov:2005} Makarov, V.~V., \&
  Kaplan, G.~H.\ 2005, \aj, 129, 2420

\bibitem[Makarov \& Robichon(2001)]{Makarov:2001} Makarov, V.~V., \&
  Robichon, N.\ 2001, \aap, 368, 873

\bibitem[Margheim(2007)]{Margheim:2007} Margheim, S. J. 2007, PhD
  thesis, Indiana Univ.

\bibitem[Mason et al.(2001)]{Mason:2001} Mason, B.~D., Wycoff, G.~L.,
  Hartkopf, W.~I., et al.\ 2001, \aj, 122, 3466

\bibitem[Mason et al.(2009)]{Mason:2009} Mason, B.\ D., Hartkopf,
  W.\ I., Gies, D.\ R.\ et al. 2009, \aj, 137, 3358

\bibitem[McGraw et al.(1974)]{McGraw:1974} McGraw, J.\ T., Dunham,
  D.\ W., Evans, D.\ S.\ et al.\ 1974, \aj, 79, 1299


\bibitem[Mermilliod et al.(1997)]{Mermilliod:1997} Mermilliod, J.-C.,
  Bratschi, P., \& Mayor, M.\ 1997, \aap, 320, 74

\bibitem[Mermilliod et al.(2009)]{Mermilliod:2009} Mermilliod, J.-C.,
  Mayor, M., \& Udry, S.\ 2009, \aap, 498, 949

\bibitem[Mermilliod et al.(1992)]{Mermilliod:1992} Mermilliod, J.-C.,
  Rosvick, J.~M., Duquennoy, A., et al.\ 1992, \aap, 265, 513

\bibitem[Moe \& Di Stefano(2017)]{Moe:2017} Moe, M., \& Di Stefano,
  R.\ 2017, \apjs, 230, 15

\bibitem[Morse et al.(1991)]{Morse:1991} Morse, J.\ A., Mathieu,
  R.\ D., \& Levine, S.\ E. 1991, \aj, 101, 1495

\bibitem[Nordstr\"om et al.(1994)]{Nordstrom:1994} Nordstr\"om, B.,
  Latham, D.\ W., Morse, J.\ A., et al.\ 1994, \aap, 287, 338

\bibitem[Ochsenbein et al.(2000)]{Ochsenbein:2000} Ochsenbein, F.,
  Bauer, P., \& Marcout, J.\ 2000, \aaps, 143, 23


\bibitem[Pearce \& Hill(1971)]{Pearce:1971} Pearce, J.\ A., \& Hill,
  G. 1971, \pasp, 83, 493

\bibitem[Pearce \& Hill(1975)]{Pearce:1975} Pearce, J.~A., \& Hill,
  G.\ 1975, PDAO, 14, 319

\bibitem[Qian \& Fan(1991)]{Qian:1991} Qian, B.-C. \& Fan,
  Q.-Y.\ 1991, Acta Astronomica Sinica, 15, 336

\bibitem[Raboud \& Mermilliod(1998)]{Raboud:1998} Raboud, D., \&
  Mermilliod, J.-C.\ 1998, \aap, 329, 101

\bibitem[Richichi et al.(1994)]{Richichi:1994} Richichi, A., Calamai,
  G., \& Leinert, Ch. 1994, \aap, 286, 829

\bibitem[Richichi et al.(1996)]{Richichi:1996} Richichi, A., Calamai,
  G., Leinert, Ch.\ et al. 1996, \aap, 309, 163

\bibitem[Richichi et al.(2002)]{Richichi:2002} Richichi, A., Calamai,
  G., \& Stecklum, B.\ 2002, \aap, 382, 178

\bibitem[Richichi et al.(2012)]{Richichi:2012} Richichi, A., Chen,
  W.~P., Cusano, F., et al.\ 2012, \aap, 541, A96

\bibitem[Rosvick et al.(1992)]{Rosvick:1992} Rosvick, J.~M.,
  Mermilliod, J.-C., \& Mayor, M.\ 1992, \aap, 255, 130


\bibitem[Rodr{\'\i}guez et al.(2000)]{Rodriguez:2000} Rodr{\'\i}guez,
  E., L{\'o}pez-Gonz{\'a}lez, M.~J., \& L{\'o}pez de Coca, P.\ 2000,
  \aaps, 144, 469

\bibitem[Royer et al.(2002)]{Royer:2002} Royer, F., Grenier, S.,
  Baylac, M.-O., et al.\ 2002, \aap, 393, 897

\bibitem[Sim{\'o}n-D{\'\i}az et al.(2017)]{Simon-Diaz:2017}
  Sim{\'o}n-D{\'\i}az, S., Godart, M., Castro, N., et al.\ 2017, \aap,
  597, A22

\bibitem[Smith \& Struve(1944)]{Smith:1944} Smith, B., \& Struve,
  O.\ 1944, \apj, 100, 360



\bibitem[Szentgyorgyi \& F\H{u}r\'esz(2007)]{Szentgyorgyi:2007}
  Szentgyorgyi, A.\ H., \& F\H{u}r\'esz, G. 2007, RMxAC, 28, 129


\bibitem[Trumpler(1921)]{Trumpler:1921} Trumpler, R.~J.\ 1921, Lick
  Observatory Bulletin, 333, 110

\bibitem[van Belle(2012)]{vanBelle:2012} van Belle, G.~T.\ 2012,
  \aapr, 20, 51

\bibitem[van Leeuwen et al.(1986)]{vanLeeuwen:1986} van Leeuwen, F.,
  Alphenaar, P., \& Brand, J.\ 1986, \aaps, 65, 309

\bibitem[Yudin(2001)]{Yudin:2001} Yudin, R.~V.\ 2001, \aap, 368, 912

\bibitem[Zorec \& Royer(2012)]{Zorec:2012} Zorec, J., \& Royer,
  F.\ 2012, \aap, 537, A120

\bibitem[Zorec et al.(2016)]{Zorec:2016} Zorec, J., Fr{\'e}mat, Y.,
  Domiciano de Souza, A., et al.\ 2016, \aap, 595, A132




\end{thebibliography}
\end{document}